# Title
Integrated and portable magnetometer based on nitrogen-vacancy ensembles in diamond

# Authors
Felix M. Stürner, Andreas Brenneis[*], Thomas Buck, Julian Kassel, Robert Rölver, Tino Fuchs, Anton Savitsky, Dieter Suter, Jens Grimmel, Stefan Hengesbach, Michael Förtsch, Kazuo Nakamura, Hitoshi Sumiya, Shinobu Onoda, Junichi Isoya, Fedor Jelezko

F. M. Stürner, Dr. A. Brenneis, T. Buck, J. Kassel, Dr. R. Rölver, Dr. T. Fuchs
Corporate Sector Research and Advance Engineering, Robert Bosch GmbH, Robert-Bosch-Campus 1, Renningen 71272, Germany
E-mail: Andreas.Brenneis@de.bosch.com

Dr. A. Savitsky, Prof. Dr. D. Suter
Fakultät Physik, Technische Universität Dortmund, Otto-Hahn-Straße 4a, Dortmund 44227, Germany

Dr. J. Grimmel, Dr. S. Hengesbach, Dr. M. Förtsch
Q.ANT GmbH, Handwerkstraße 29, Stuttgart 70565, Germany

Dr. K. Nakamura
Leading-Edge Energy System Research Institute, Fundamental Technology Dept., Tokyo Gas Co., Ltd., Yokohama 230-0045, Japan

Dr. H. Sumiya
Advanced Materials Laboratory, Sumitomo Electric Industries, Ltd., Itami 664-0016, Japan

Dr. S. Onoda
Takasaki Advanced Radiation Research Institute, National Institutes for Quantum and Radiological Science and Technology, Takasaki 370-1292, Japan

Prof. Dr. J. Isoya
Faculty of Pure and Applied Sciences, University of Tsukuba, Tsukuba 305-8573, Japan

F. M. Stürner, Prof. Dr. F. Jelezko
Institute for Quantum Optics and Center for Integrated Quantum Science and Technology (IQ$^{ST}$), Ulm University, Albert-Einstein-Allee 11, Ulm 89081, Germany

# Abstract

Magnetic field sensors that exploit quantum effects have shown that they can outperform classical sensors in terms of sensitivity enabling a range of novel applications in future, such as a brain machine interface. Negatively charged nitrogen-vacancy (NV) centers in diamond have emerged as a promising high sensitivity platform for measuring magnetic fields at room temperature. Transferring this technology from laboratory setups into products and applications, the total size of the sensor, the overall power consumption, and the costs need to be reduced and optimized. Here, we demonstrate a fiber-based NV magnetometer featuring a complete integration of all functional components without using any bulky laboratory equipment. This integrated prototype allows portable measurement of magnetic fields with a sensitivity of $344\,\mathrm{pT\,Hz^{-1/2}}$.

## 1. Introduction

Sensing and metrology by means of adapting quantum technologies have attracted a high degree of attention in recent years, as evidenced by numerous large publicly funded programs worldwide and a growing interest of companies in this field of technology. A very promising quantum sensing approach, which represents the transition from basic research to product application, is based on the negatively charged nitrogen-vacancy (NV) centers in diamond. These color centers comprise high magnetic sensitivity at ambient conditions with large dynamic range and high spatial resolution. [1] [2] In addition, this technology can be applied to measure temperature, [3] electric fields, [4] and pressure. [5]

The detection of magnetic field utilizes the Zeeman effect seen by the NV center's electron spin, whose spin state can be optically initialized, manipulated using microwave fields, and optically read out with long coherence times. [6] Limitations on the achievable sensor performance for NV ensembles arise from several factors, which are interconnected and mutually dependent, and impede a straightforward optimization of the performance. The sensitivity depends on the quality of the diamond material with parameters of NV concentration, NV orientation, NV charge state, isotopic compositions, concentration of paramagnetic impurities, and the material parameters, such as strain limiting the coherence properties of the spins. [7] For the optical initialization of the NV spins, the optical power, wavelength, polarization, and homogeneity of the excitation light need to be optimized for a high sensitivity NV ensemble magnetometer. [8] Further effort is necessary to provide a homogeneous offset magnetic field in addition to a sufficiently strong and homogeneous microwave field [9]. Due to diamond's high refractive index, solutions need to be developed to efficiently extract the fluorescence light that is emitted by the NV centers. [10] Furthermore, the implementation of pulsed measurement protocols can improve the sensitivity by decreasing the impact of noise sources, *e.g.* decoherence due to paramagnetic spins or temperature fluctuations. [11] [12]

Recent achievements in NV diamond magnetometry have shown detection of AC fields with a sensitivity of $0.9\,\mathrm{pT\,Hz^{-1/2}}$, [2] recording of neuronal action potentials with a sensitivity of $15\,\mathrm{pT\,Hz^{-1/2}}$, [11] and a sensitivity of $0.9\,\mathrm{pT\,Hz^{-1/2}}$ in the low frequency range. [13] Vector magnetometry experiments have been demonstrated for DC [14] and AC fields [15] enabling NV centers for a wide range of applications. Potential applications could be implemented in the field of medical diagnostics or brain-machine interface. [16] In addition to high sensitivity, NV centers magnetometers can cover high dynamic range. However, most of these impressive results have been achieved on bulky setups with constraints to a laboratory environment. For a mobile application of a NV-based magnetometer it is indispensable to integrate and miniaturize the sensor components to a compact and portable device. Based on different approaches, recent works have focused on the development of prototypes that can enable the NV centers for industrial product applications. [17] [18] [19] [20] [21] [22] Here, we demonstrate a fiber-integrated magnetometer based on NV centers in diamond. The compact design of the magnetic field sensor was achieved by using a single-mode fiber for the optical initialization of the NV centers and deploying a balanced detection scheme built up by two photodiodes positioned close to the diamond. Furthermore, our integrated prototype comprises all necessary components, such as offset magnetic field, laser source, microwave generator, and signal processing unit. This portable setup yields a sensitivity of $\approx 344\,\mathrm{pT\,Hz^{-1/2}}$.

## 2. Materials and methods

The NV center is a color center in the lattice structure of a diamond consisting of a nitrogen atom and an adjacent vacancy spot, see **Figure 1**. The NV centers are susceptible to magnetic fields through the Zeeman effect yielding a magnetic field dependent electron spin resonance for the spin transitions from the $m_s = 0$ to the $m_s = \pm 1$ ground states, see Figure 1b. Zeeman splitting leads to a shift of the resonance frequencies $f_\pm$ of the spin transitions of NV centers as [23]

$$f_\pm \approx D_{gs} + \beta \cdot \Delta T \pm \gamma_{NV} B_0, \tag{1}$$

where $D_{gs} \approx 2.87\,\mathrm{GHz}$ is the zero-field splitting due to the spin-spin interaction at room temperature, $\beta = dD_{gs}/dT \approx -75.0\,\mathrm{kHz\,K^{-1}}$ the temperature dependence of the zero-field splitting, [23] $\Delta T$ the temperature shift from a room temperature condition, $\gamma_{NV} = \mu_B g_s/h \approx 28.024\,\mathrm{GHz\,T^{-1}}$ the gyromagnetic ratio with $\mu_B$ the Bohr magneton, $g_s \approx 2.0028$ the approximately isotropic g-factor, h the Planck constant, and $B_0$ the projection of the magnetic field on the respective NV axis. [24] As the NV spin states can be both optically initialized and read out, optical excitation in combination with a resonant electron spin excitation leads to a reduced fluorescence intensity. Due to the Zeeman effect, the shift of the resonance frequencies from $D_{gs}$ is proportional to the magnetic field.

The optically detected magnetic resonance (ODMR), as a standard technique, can be performed either in a continuous wave (CW) or pulsed mode. [25] The CW ODMR measurement is characterized by a simultaneous application of excitation light and microwave field to initialize and manipulate the NV spins. It is typically easier to implement compared to the pulsed scheme.

The sensor head of the integrated NV magnetometer is shown in **Figure 2** and the complete experimental setup can be seen in **Figure S1.** The integrated sensor head includes a diamond with NV centers, which is connected to a lens and an optical single mode fiber, a microwave (MW) circuit for providing an oscillating magnetic field necessary for the spin manipulation, and a printed circuit board (PCB) containing two photodiodes for a balanced detection scheme of the emitted fluorescence and the excitation laser light. Microwave source and laser were located externally and connected to the sensor head by radiofrequency cables and an optical fiber, respectively.

We used a fiber coupled, compact laser module from Q.ANT GmbH with an emitting wavelength of $521.9$ nm (dimensions of the laser module: $35 \times 110 \times 115$ mm³) providing $23.5$ mW at the end of the fiber. This laser is coupled to a polarization-maintaining (PM) single mode optical fiber (Thorlabs PM-S405-XP, core diameter $3.0\,\mu$m, $NA = 0.12$). A direct connection between fiber and diamond would result in an inhomogeneous optical excitation of the NV centers along the optical path, see **Figure S2**a. We used therefore a gradient refractive index (GRIN) rod lens (GRINTECH GT-LFRL-050-025-50-NC (670), $NA = 0.5$) to collimate the light that exits the fiber before it enters the diamond, see FigureS9b. The GRIN rod lens and fiber were connected by an epoxy glue and inserted into a pigtailed glass ferrule to stabilize the assembly. The diamond was directly glued to the surface of the GRIN lens with a droplet of polydimethylsiloxane (PDMS).

We used a (111)-oriented plate with dimensions of 0.8 × 0.8 × 0.5 mm³ obtained by laser-cutting and polishing from a 99.97 % $^{12}$C enriched diamond single crystal grown by the temperature gradient method at high pressure and high temperature (HPHT) conditions. The original crystal was irradiated with 2 MeV electrons ($2 \times 10^{18}$ cm$^{-2}$) at room temperature and annealed at 1000 C for 2 h in vacuum resulting in the concentrations of [NV$^-$] ∼ 0.4 ppm and [P1] ∼ 2.0 ppm measured by electron spin resonance (ESR).

For the microwave-based spin manipulation, we used a double split-ring resonator (**Figure S9**) formed by two inductively coupled transmission lines, which were terminated by two capacitive gaps, in order to provide a microwave field of sufficient amplitude. [26] The resonator was fabricated on a 640 µm thick Rogers RO3010™ substrate with a copper metallization of 20 µm on both sides (PCB dimensions of 15 × 24.5 × 0.72 mm³). An MMCX connector was used to supply the resonator with microwave power, transferred by a 50 Ω microstrip line coupled to the double split-ring resonator. The bandwidth of the resonator is 23.48 ± 0.08 MHz and the resonance frequency can be tuned in the range from 2.8 to 3.0 GHz by putting a metallized plate (resonator tuning stick, Figure 2) on the resonator structure. Further details of the microwave resonator will be given in Section 10 of the supplementary information.

Microwaves in the GHz frequency range were delivered by a compact local oscillator (Windfreak Synth USB II) with an output power of 3 dBm and a fixed frequency. Additionally, we used a second source (RedPitaya STEM 125-14 V1.0) to synthesize the baseband in the MHz frequency range. The baseband signal was pre-amplified using a low noise amplifier (Mini-Circuits ZFL-1000LN+). Both signals were subsequently mixed using a low conversion loss mixer (Mini-Circuits ZFM-4212+) and amplified using three low noise microwave amplifiers (Mini-Circuits ZX60-P33ULM+) with a gain of 8.8 dB each (at a frequency of 3.0 GHz). The frequency of the baseband signal was modulated with a modulation frequency $f_{\mathrm{mod}}$ and a modulation depth $f_{\mathrm{depth}}$, *i.e.* $f_{\mathrm{BB}}(t) = f_{\mathrm{BB}} + f_{\mathrm{depth}} \cdot \sin(2\pi f_{\mathrm{mod}} t)$, where $f_{\mathrm{BB}}$ is the mean frequency of the baseband. This signal was mixed with the fixed frequency $f_{LO}$ of the local oscillator resulting in the overall mean frequency $f_c = f_{LO} + f_{BB}$. [27] The following analysis will refer to $f_c$. For an enhancement of the magnetic sensitivity we simultaneously applied three mean frequencies $f_c$, $f_c - f_{HFS}$, and $f_c + f_{HFS}$ in order to address all three [14]N hyperfine resonances with a frequency splitting $f_{HFS}$ of 2.16 MHz simultaneously. [18] [28] In the following, we will label such measurements with '3 HFS on', while '3 HFS off' means that only the mean frequency $f_c$ is applied. If not explicitly mentioned, the default settings for the presented results is '3 HFS on' and the given values will refer to the total power of the three frequencies.

The fluorescence emitted by the diamond was collected by a photodiode (PD) located on the diamond surface opposite to the fiber and GRIN lens. This is in contrast to previous publications, where the fiber was used both for optical excitation and fluorescence collection. [29] [30] [31] [32] For this purpose, we used a chip photodiode (Hamamatsu S12915-33CHIP) with a photosensitive area of $2.4 \times 2.4$ mm² and a responsivity of $R_{670nm} \approx 0.54$ A W$^{-1}$ (at 670 nm wavelength). A 1.45 µm thin distributed Bragg reflector (DBR) grown by chemical vapor deposition of several layers of amorphous silicon carbide (aSiC, $n \approx 2.6$) and silicon oxide (SiO$_2$, $n \approx 1.5$) was deposited on the surface of the photodiode acting as a long-pass filter to block the laser excitation light. [17] Besides the sole detection of the emitted fluorescence of the NV centers, we additionally monitored the laser intensity by a second photodiode with responsivity of $R_{522nm} \approx 0.40$ A W$^{-1}$. This photodiode was positioned next to the first photodiode on the common PCB and was coated with a customized DBR blocking the fluorescence light (short-pass filter). Both photodiodes were reverse biased with a voltage of 6 V. Implementing long cut holes on the photodetection board enabled a variable alignment of the photodiodes relative to the position of the diamond, such that the ratio of detected fluorescence and laser (reference) signal can be adjusted. Here, our design represents a compromise between full detection of the fluorescence signal and maintaining the ability of balanced detection. For this purpose, the signals of the two photodiodes were fed to a logarithmic transimpedance amplifier (Texas Instruments LOG114), whose pre-amplified output was digitized and demodulated (according to the applied modulation frequency) by the analogue-to-digital module of the RedPitaya ( 14 bit, 125 MSamples/s). The demodulated signal $S_{\text{demod}}$, shown in **Figure 3**a, indicates the peak-to-peak amplitude of a signal oscillating at the carrier frequency $f_{\text{mod}}$.

For the experiments, we used the technique of ODMR to measure magnetic fields. Therefore, the demodulated signal output $S_{\text{demod}}$ of the RedPitaya resembles to the derivative of the typical Lorentzian profile of the ODMR spectrum. Close to the resonance frequency, $S_{\text{demod}}$ is proportional to the detuning $\Delta f$ from the corresponding resonance frequency $f_{\pm}$ (Figure 3a). We extracted information about the linewidth $\Delta v$ and contrast $C$ of the electron spin resonance by integrating the demodulated signal $S_{\text{demod}}$ and fitting the resulting data $S_{\text{integ}}$ to a Lorentzian function $g(x)$ of

$$g(x) = A \frac{(\Delta v/2)^2}{(x - x_0)^2 + (\Delta v/2)^2}, \qquad (2)$$

where $A$ is an amplitude factor linked to spin contrast $C$, $\Delta v$ is the full width at half maximum of the resonance curve, and $x_0$ is the resonance frequency (see Figure 3b). Details of the signal processing, including the signal integration and the determination of the contrast, are summarized in Section S4 of the supplementary information.

In order to spectrally separate the four NV orientations and hyperfine splittings, we used a self-constructed, current carrying Helmholtz coil, is a necessary functional part of the sensor head. The Helmholtz coil consists of 35 windings on each side providing an offset magnetic field of $\approx 1.07 \, \text{mT}$ at a current of 0.5 A. The windings were designed in a tapered shape to fulfill the Helmholtz coils' condition, *i.e.* the distance between the two coils equals the radius of the coils, resulting in field deviations of less than 0.1 % inside the volume of the diamond, see **Figure S3**.

A diameter of 22.5 mm for the smallest winding and a diameter of 36.2 mm for the largest winding was used.

## 3. Experimental results

An ODMR spectrum of the NV ensembles recorded with an offset magnetic field generated by the tapered Helmholtz coils is shown in **Figure 4**a. The measured demodulated signal $S_{\text{demod}}$ shows the spin transitions from the $m_s = 0$ to $m_s = 1$ states including the three sublevels that result from the hyperfine interaction with the nuclear spin of the $^{14}$N isotope. All four possible orientations contribute to the spectrum depicted in Figure 4a. The color-coded arrows are related to the electron spin resonances of the four NV orientations. Due to the simultaneous manipulation of the three hyperfine splittings, five characteristic zero-crossings can be observed in Figure 4a for each NV axis. In order to show the increase in signal amplitude due to the simultaneous hyperfine spin manipulation in comparison to a standard spin manipulation, we plotted the spectra in a common plot superimposed, see Figure 4b. The simultaneous excitation of the three $^{14}$N hyperfine split transitions should result in a threefold increase of the contrast. However, we only find a factor of approximately 2.61 larger amplitude, which may be due to an uneven power level of the combined three mean microwave frequencies, *i.e.* $f_c$, $f_c - f_{\text{HFS}}$, and $f_c + f_{\text{HFS}}$, see **Figure S5**a.

In the following, we evaluated the optimal working point of our NV sensor in order to improve the sensitivity $\eta_B$ by varying the adjustable parameters, *i.e.* modulation depth $f_{\text{depth}}$, microwave power $P_{\text{MW}}$, integration time $t_{\text{int}}$, and modulation frequency $f_{\text{mod}}$, while the laser power was fixed. The constant power of the laser module of 23.5 mW corresponds to an estimated average optical intensity of $I_{\text{opt}} \approx 390 \text{ W cm}^{-1}$ along the optical path inside the diamond (for further details see Section 10 in the supplementary information). Based on the results by Dréau *et al.* and Levine *et al.*, this value of the optical intensity corresponds to the excitation regime where the tradeoff between high contrast and small linewidth results in an optimal value for the sensitivity. [9] [12] For the following analysis, we focused on the NV orientation with the largest field projection. As depicted in Figure 4a, the mean frequency $f_c$ was swept to record an ODMR spectrum of the demodulated signal $S_\text{demod}$. We applied a polynomial fit of 6[th] order to the measured data of $S_\text{demod}$ to evaluate the linear slope $m$ at the zero-crossing as well as the noise. The noise was calculated as the standard deviation $\sigma$ of the residuum $\epsilon$ that results from subtracting the polynomial fit from the measured data, see Figure 3a. Using these quantities and the gyromagnetic ratio $\gamma_\text{NV} \approx 28.024\,\text{GHz}\,\text{T}^{-1}$, [24] we approximated the sensitivity for a measurement time $t_\text{int}$ in the linear range close to the zero-crossing point of $S_\text{demod}$ with

$$\eta_B = \frac{\sigma}{\gamma_\text{NV} \cdot m} \sqrt{t_\text{int}}. \tag{3}$$

The results of this optimization analysis are summarized in **Figure 5**. As we will present in the following, the best value of this analysis yields $\eta_B \approx 424\,\text{pT}\,\text{Hz}^{-1/2}$ (without magnetic shielding).

First, we studied the dependence of the modulation depth $f_\text{depth}$ on the sensitivity, as shown in Figure 5a. El-Ella *et al.* found an optimal modulation depth, which maximizes the measured slope $m$, corresponding to half of the resonance linewidth. [33] The extracted slope reached a maximum at $40\,\text{kHz}$ modulation depth and the noise showed no clear trend on the modulation depths, see **Figure S8**c. Fitting the data of $S_\text{integ}$ with the Lorentzian function from Equation (2) a linewidth of $81.8 \pm 2.4\,\text{kHz}$ and a contrast of $0.43 \pm 0.01\,\%$ were extracted for $f_\text{depth} = 40\,\text{kHz}$ (Figure S8b), which is in good agreement to the findings of El-Ella *et al.*. Therefore, $f_\text{depth} = 40\,\text{kHz}$ was chosen for the following measurements.

In a second step, the influence of the applied microwave power was evaluated. We observed an increase of the measured linewidth for microwave powers larger than 0.1 mW (Figure S9b) due to power broadening. [9] For microwave powers below 0.1 mW, the linewidth seems to be unaffected by power broadening. We extracted an inhomogeneously broadened linewidth $\Delta\nu_{\text{inh}}$ of approximately 76.47 kHz by applying a theoretical model according to Jensen *et al.* for the linewidth of NV ensembles to the measured data, see Figure S9b. [34] Regarding the measured contrast that results from the simultaneous excitation of the three hyperfine levels, a maximum value of $\approx 1.0$ % is achieved for $P_{\text{MW}} \approx 1.45$ mW. However, for applied powers above 1.45 mW the contrast deteriorated as microwave induced power broadening dominated over the optical pumping rate and became major linewidth determining factor. We found that the slope was largest for a moderate microwave power of $\approx 58$ µW. The noise had a similar dependence on the microwave power. As a result, the overall sensitivity showed a rather weak dependence on the microwave power in the range from 0.02 mW to 0.2 mW.

In the third step, we varied the integration time based on the optimal parameters to date to record slope and noise. The best value for the sensitivity was obtained for an integration time of 20 ms, see Figure 5c, which is linked to the 50 Hz main frequency of the European power lines. The measured values of slope and noise are shown in **Figure S10**a.

In the fourth step, the modulation frequency was varied under the parameters determined so far, see Figure 5d. We found a maximum slope for $\approx 1$ kHz modulation frequency, see Figure S7b. While the slope decreased for $f_{\text{mod}} > 1$ kHz, the noise level was almost constant for all investigated values of $f_{\text{mod}}$, see **Figure S11**c. Modulation frequencies smaller than 1 kHz resulted in a decreased slope, which can be understood by the used high-pass filter in our photodetection circuit, see **Figure S6** and **Figure S7**b. The measured values of linewidth and contrast of the modulation frequency variation are summarized in Figure S11b. The contrast resembled the behavior of the slope and decreased as the modulation frequency was increased, see Figure S11a. A maximum contrast of 0.62 % was obtained for $f_{\text{mod}} = 1\,\text{kHz}$. The linewidth showed a slight increase up to a modulation frequency of 10 kHz; for larger $f_{\text{mod}}$ the linewidth broadened significantly. Here, we note that the resulting shape of the measured signal significantly deviated from a Lorentzian profile for $f_{\text{mod}} > 16\,\text{kHz}$ such that the linewidth and contrast of the electron spin resonance cannot be extracted correctly. We will refer to this behavior in the discussion section.

Based on the described method of slope and noise characterization the best measured sensitivity was $\approx 424\,\text{pT}\,\text{Hz}^{-1/2}$, which was achieved for a frequency modulation and simultaneous excitation of the three hyperfine interactions with $f_{\text{depth}} = 40\,\text{kHz}$, $P_{\text{MW}} = 58\,\mu\text{W}$, $t_{\text{int}} = 20\,\text{ms}$, and $f_{\text{mod}} = 1\,\text{kHz}$, see Figure 5d.

The magnetic sensitivity of the sensor device was further evaluated using a magnetic shielding. For this purpose, we put a 180 mm diameter end cap of a mu-metal shield (Twinleaf MS-2) on top of the sensor and characterized the magnetic sensitivity with the method described above. We measured a sensitivity of $369\,\text{pT}\,\text{Hz}^{-1/2}$ for this partial enclosure of the NV sensor. A further reduction of the measured sensitivity was achieved by a complete enclosure of the NV sensor using two shielding end caps resulting in a sensitivity to $344\,\text{pT}\,\text{Hz}^{-1/2}$. This is an improvement of factor 1.23 compared to the unshielded setting.

Besides the determination of the magnetic sensitivity by taking the measured parameters of slope and noise of the demodulated signal at the zero-crossing point, we further evaluated the sensor performance under the application of a test magnetic field. A copper wire was wound around the pair of Helmholtz coils, on which a signal with square-wave function of 2 Hz frequency and amplitude of 2.18 mA was applied by an arbitrary waveform generator (Agilent 33220A). The applied signal corresponds to an induced magnetic field of $\approx 333\,\text{nT}$ with a field orientation parallel to the applied offset magnetic field; a detailed description of the generated test field is summarized in Section 11 of the supplementary information. We analyzed the sensitivity by recording the demodulated signal as a time trace under the application of the test field, see **Figure 6**a. Each data point of the recorded time trace corresponds to an integration time of $t_{\text{int}} = 20$ ms. The noise on a single 1 s step was evaluated by fitting the histogram of the recorded data set with a Gaussian distribution. Indicating a standard deviation $\sigma$ of $\approx 3.03$ nT, this measurement extrapolated to a sensitivity of $\sigma \cdot \sqrt{t_{\text{int}}} \approx 429$ pT Hz$^{-1/2}$.

The sensor performance was further analyzed by recording time traces of magnetic stimuli, *i.e.* the microwave frequency was set to the zero-crossing, and by calculating the Allan deviation $\sigma_A$ of this data. In general, the Allan deviation determines the deviation of the measured data binned to intervals with averaging length $\tau$. [35] In contrast to the standard deviation, the Allan deviation analyzes data sets depending on the averaging time $\tau$ and thus providing time dependent noise information. The Allan variance $\sigma_A^2(\tau)$ is calculated by [2]

$$\sigma_A^2(\tau) = \langle (\bar{y}_{n+1} - \bar{y}_n)^2 \rangle, \tag{4}$$

where $\bar{y}_n$ and $\bar{y}_{n+1}$ are the mean values of two consecutive intervals of length $\tau$. The calculated values of the Allan deviation were adjusted by the slope $m$ to translate deviations in the measured signal into magnetic field deviations, see Figure 6b. We additionally enclosed the NV sensor using the magnetic shielding of two mu-metal half-shelves, as described above, to evaluate the influence of surrounding magnetic noise sources on the achievable sensor. The noise of a magnetically insensitive case was additionally measured without using any magnetic shielding. For this purpose, we applied a microwave frequency with 19.45 MHz detuning from the selected resonance frequency.

For short averaging times $\tau$ in the range of $1 - 100$ ms, the mean sensitivity level of the magnetically sensitive measurement was improved from 440 pT Hz$^{-1/2}$ to 396 pT Hz$^{-1/2}$ through introducing the shielding. Considering a 20 ms integration time derived by the optimization steps, the unshielded measurements revealed a sensitivity of $436 \, \text{pT} \, \text{Hz}^{-1/2}$ and $365 \, \text{pT} \, \text{Hz}^{-1/2}$ was achieved for the shielded measurement environment. The shielded measurements showed a flat behavior for $1 - 100$ ms comparable to the magnetically insensitive measurement, whereas the sensor in the unshielded environment varied in its sensitivity result. An increase in the measured sensitivity for both settings was observed for time scales larger than 100 ms. Here, we note that we observed fluctuations in the data from measurement to measurement for the unshielded setting, which may be due to magnetic fields that prevail in the background.

The logarithmic balancing of the two photodiodes prevented noise measurements of the detection output without laser and microwave excitation such that the contribution of electronic noise to the overall sensitivity cannot be solely analyzed. A noise floor of $\approx 364 \, \text{pT} \, \text{Hz}^{-1/2}$ was characterized for 1 ms integration time recorded by the magnetically insensitive measurement, see Figure 6b, indicating a lower limit of the achievable sensitivity of our device.

As an application, we used the NV sensor to measure the change in magnetic field $\Delta B$ due to the movement of an elevator, see **Figure 7**. For this purpose, we positioned the NV sensor roughly 4 m away from the doors of an elevator in the third floor. The NV magnetometer recorded the movement of the elevator by tracking the resonance frequency of the selected spin transition using a feedback loop to adjust the applied microwave frequency. From the measured change in magnetic field, shown in Figure 7, it can be clearly identified when the elevator started with the upward movement (observed by an increase of the field change), opened its doors (by an additional increase of the field change compared to the "standing" case at the third level), and waiting period in this upper level. The closing of the doors, the downward movement, as well as the repeated opening of the doors were detected by the sensor in the form of a decreasing change decreasing $\Delta B$ compared to the starting situation.

## 4. Discussion and outlook

For the evaluation of the achieved sensor performance, it is essential to have a look on its fundamental limit. Due to the optical readout of the spin states, the NV magnetometer is limited by photon shot-noise and is given by [10]

$$\eta_{PSN} = P_F \cdot \frac{1}{\gamma_{\text{NV}}} \cdot \frac{\Delta \nu}{C} \cdot \frac{1}{\sqrt{R}}, \tag{5}$$

where $P_F$ is a numerical factor resulting from the shape of the resonance with $P_F = 4/(3\sqrt{3}) \approx 0.77$ for a Lorentzian profile, $R$ is the detected photon rate. We determined the photon rate from the average photocurrent $I_{\text{PD}}$ by measuring the voltage $U_{\text{shunt}}$ drop at a $R_{\text{shunt}} = 10\,\Omega$ shunt resistor. For a measured $U_{\text{shunt}}$ of 0.40 mV the photon rate $R = \frac{I_{\text{PD}}}{e} = \frac{U_{\text{shunt}}}{R_{\text{shunt}} \cdot e}$ results in $\approx 2.47 \cdot 10^{14}$ Hz, with $e$ being the elementary charge. Using Equation (5) and the measured linewidth of 92.19 kHz and contrast of 0.62 % the photon shot-noise limited sensitivity is $\approx 26\,\text{pT}\,\text{Hz}^{-1/2}$. This value is a factor of $\approx 13$ smaller than the best measured sensitivity value of $\approx 344\,\text{pT}\,\text{Hz}^{-1/2}$. Regarding this discrepancy a factor of $\sqrt{2} \approx 1.41$ can be explained by the usage of the balanced photodetection unit based on two single independent photodiodes compared to a single photodiode usage. Further deviations may arise due to temperature fluctuations, which are not included in the calculation of the photon shot-noise limit, but may worsen the recorded sensitivity in terms of a broadened linewidth. Besides these effects, we considered the deviation from the photon shot noise limit to arise from fluctuations of the laser and its interplay with the limited common mode rejection of the balanced photodetector module.

Finding the optimal working point of the NV magnetometer by optimizing the relevant operation parameters is essential to obtain a highly sensitive device at efficient usage of the sensor power supply. Due to the interplay between optical and microwave excitation it may be possible that altering the applied laser power improves the sensitivity. We further note that the parameter evaluation of the modulation frequency was not straight forward compared to the parameter sweeps of microwave power and modulation depth. This is mainly linked to the implemented high-pass filter in the transimpedance amplification circuit of the balanced detection unit. Based on simulations of the electronic circuit using LTspice®, we expected the output of the logarithmic transimpedance amplification circuit to show a flat behavior over a frequency range between $1 - 50$ kHz, see Figure S6. However, experimental results of the contrast and slope determination revealed a maximum value at $1$ kHz modulation frequency, as depicted in Figure S7b. The measured signal deteriorated for higher modulation frequencies, which can be explained by a relatively small optical pumping rate compared to the modulation frequency of the microwave. We estimated the pumping rate to be $\approx 32$ kHz (see Section 10 of the supplementary information), so that under these conditions the pumping of the NV system is too slow to follow the fast modulation resulting in a reduced signal. In the simulation the dynamics of the NV centers was neglected. Therefore, we assume that the discrepancy between experimental observation and simulative prediction results from the complex excitation dynamics of the NV centers.

The method of applying an additional magnetic test field to the NV sensor and measuring the noise on the demodulated signal at the point of zero-crossing, *i.e.* resonance condition, confirmed the sensitivity that was obtained from the analysis of the slope and its residuum, see Figure 3a. For short averaging times in the range of $1 - 100$ ms the achievable sensitivity varied, which may be due to (magnetic) noise/fluctuations in the sensor's environment. The analysis of the Allan deviation revealed that the performance worsened for sensing times exceeding 1 s independent on the sensor's environment, *i.e.* unshielded and magnetically shielded. Observed drifts in the recorded time traces may be linked to temperature variations and need to be mitigated to improve both the sensor's sensitivity and stability on larger time scales. Recent publications have shown how to set up a sensor with temperature invariance by simultaneously addressing both transitions from the $m_s = 0$ to $m_s = \pm 1$, *i.e.* making use of double quantum magnetometry. [23] [36] A complete enclosure of the NV magnetometer in a magnetic shielding showed an improvement of the sensitivity.

The fluorescence collection efficiency could be further optimized, as we detected less than one percent of the power that was used for excitation (a calculated fluorescence power of 73.1 µW based on the detected photocurrent for a fluorescence wavelength of 670 nm and an estimated excitation output power at the GRIN lens of 18.3 mW). This low detection efficiency was mainly linked to the position of the two photodiodes, which were used for the detection of laser and fluorescence light, relative to the diamond and fiber. An instant improvement in the fluorescence detection efficiency could be achieved by avoiding the step in refractive index between the diamond and the photodiode by filling the gap with a material having a larger refractive index than air. This could mitigate the problem of total internal reflections of the emitted fluorescence due to diamond's high refractive index. The DBR filters deposited on the surface of the photodiode further reduced the detectable fluorescence intensity as these filters showed a transmission of $\approx 78\,\%$.

A further improvement of sensor performance could be achieved by equalizing the microwave power levels at the three hyperfine frequencies, since different power levels prevent a full exploitation of simultaneously addressing hyperfine levels.

Besides the implementation of temperature invariance of the NV sensor, another possibility to enhance the achievable sensitivity could be the use of flux concentrators, as recently shown by Fescenko *et al.*, which collect the magnetic flux from a large area and direct it to the NV sensor. [13] The use of such concentrators must nevertheless be viewed critically from the point of view of the intended application, as any restrictions on the use of the sensor could be associated with the use of these concentrators, *e.g.* for a vector magnetometry application. Setting up the sensor device as a vector magnetometer, we would need to implement a simultaneous addressing of at least three NV orientations in the microwave and signal processing circuitry. [15] Tracking fast changing magnetic signals would further require a closed loop control for tracking the NV resonance frequencies. Further miniaturization for mobile applications of NV magnetometers could take advantage of using complementary metal-oxide-semiconductors (CMOS) as shown by Kim *et al.*. [21]

Regarding power consumption of the NV magnetometer, the power efficiency of the microwave resonator could be further enhanced by optimizing the coupling property, *e.g.* by placing additional dielectric material on the coupling gap. This could reduce the need of amplification or the input power for the spin manipulation.

The compactness and portability of our NV magnetometer was designed to pave the way for a wide range of applications including GPS-free navigation on earth and space missions as well as the detection of biomagnetic signals, *e.g.* for magnetocardiography and magnetoencephalography. [16] By implementing the above mentioned measures, the sensor's sensitivity could be improved to the required level of $\approx 10\,\text{pT}\,\text{Hz}^{-1/2}$ for magnetocardiography. In addition, our NV magnetometer could be used for the detection of electric currents, *e.g.* for battery diagnostics.

## 5. Conclusion

In conclusion, we have demonstrated an integrated portable prototype of a magnetometer based on the NV centers in diamond. The diamond attached to an optical fiber, the microwave resonator, and a balanced detection scheme built up by two photodiodes were co-integrated into a compact sensor head. Our focus in setting up this prototype was put on the size and compactness of the complete sensor device while enabling high sensitivity measurements. This includes the choice of laser source, microwave equipment, and signal processing such that the sensor device can be used in real world applications outside laboratory environments. Using the frequency modulation technique for measuring electron spin resonance we achieved a magnetic field sensitivity of $344\,\text{pT}\,\text{Hz}^{-1/2}$. We further discussed solutions to improve the sensor performance and abilities to increase the integration level.

**Acknowledgements**
The authors thank Jens Baringhaus for the support of scanning electron microscope images of the resonator structure.
We acknowledge financial support by the German Federal Ministry of Education and Research (BMBF) within the "BrainQSens" project (No. 13N14435) and the "MiLiQuant" project (No. 13N15062). This work was also financially supported by the European Quantum Technology Flagship Horizon 2020 through the "ASTERIQS" program (No. 820394). In addition, we acknowledge the support of the Japan Society for the Promotion of Science (JSPS) KAKENHI (No. 20H00340).

**Conflict of interests**
F. M. Stürner, A. Brenneis, T. Buck, J. Kassel, R. Rölver, and T. Fuchs are co-inventors of a pending patent application related to this work. The remaining authors declare no competing financial interests.

**References**

[1] J. M. Taylor, P. Cappellaro, L. Childress, L. Jiang, D. Budker, P. R. Hemmer, A. Yacoby, R. Walsworth and M. D. Lukin, "High-sensitivity diamond magnetometer with nanoscale resolution," *Nature Physics,* vol. 4, pp. 810-816, 2008.

[2] T. Wolf, P. Neumann, K. Nakamura, H. Sumiya, T. Ohshima, J. Isoya and J. Wrachtrup, "Subpicotesla Diamond Magnetometry," *Physical Review X,* vol. 5, no. 4, p. 041001, 2015.

[3] P. Neumann, I. Jakobi, F. Dolde, C. Burk, R. Reuter, G. Waldherr, J. Honert, T. Wolf, A. Brunner, J. H. Shim, D. Suter, H. Sumiya, J. Isoya and J. Wrachtrup, "High-Precision Nanoscale Temperature Sensing Using Single Defects in Diamond," *Nano Letters,* vol. 13, pp. 2738-2742, 2013.

[4] F. Dolde, H. Fedder, M. W. Doherty, T. Nöbauer, F. Rempp, G. Balasubramanian, T. Wolf, F. Reinhard, L. C. L. Hollenberg, F. Jelezko and others, "Electric-field sensing using single diamond spins," *Nature Physics,* vol. 7, p. 459, 2011.

[5] M. W. Doherty, V. V. Struzhkin, D. A. Simpson, L. P. McGuinness, Y. Meng, A. Stacey, T. J. Karle, R. J. Hemley, N. B. Manson, L. C. L. Hollenberg and S. Prawer, "Electronic Properties and Metrology Applications of the Diamond NV^- Center under Pressure," *Physical Review Letters,* vol. 112, no. 4, p. 047601, 2014.

[6]  G. Balasubramanian, P. Neumann, D. Twitchen, M. Markham, R. Kolesov, N. Mizuochi, J. Isoya, J. Achard, J. Beck, J. Tissler, V. Jacques, P. R. Hemmer, F. Jelezko and J. Wrachtrup, "Ultralong spin coherence time in isotopically engineered diamond," *Nature Materials,* vol. 8, pp. 383-387, 2009.

[7]  J. Achard, V. Jacques and A. Tallaire, "Chemical vapour deposition diamond single crystals with nitrogen-vacancy centres: a review of material synthesis and technology for quantum sensing applications," *Journal of Physics D: Applied Physics,* vol. 53, p. 313001, 2020.

[8]  K. Beha, A. Batalov, N. B. Manson, R. Bratschitsch and A. Leitenstorfer, "Optimum Photoluminescence Excitation and Recharging Cycle of Single Nitrogen-Vacancy Centers in Ultrapure Diamond," *Physical Review Letters,* vol. 109, no. 9, p. 097404, 2012.

[9]  A. Dréau, M. Lesik, L. Rondin, P. Spinicelli, O. Arcizet, J.-F. Roch and V. Jacques, "Avoiding power broadening in optically detected magnetic resonance of single NV defects for enhanced dc magnetic field sensitivity," *Physical Review B,* vol. 84, no. 19, p. 195204, 2011.

[10] J. F. Barry, J. M. Schloss, E. Bauch, M. J. Turner, C. A. Hart, L. M. Pham and R. L. Walsworth, "Sensitivity optimization for NV-diamond magnetometry," *Reviews of Modern Physics,* vol. 92, no. 1, p. 015004, 2020.

[11] J. F. Barry, M. J. Turner, J. M. Schloss, D. R. Glenn, Y. Song, M. D. Lukin, H. Park and R. L. Walsworth, "Optical magnetic detection of single-neuron action potentials using quantum defects in diamond," *Proceedings of the National Academy of Sciences,* 2016.

[12] E. V. Levine, M. J. Turner, P. Kehayias, C. A. Hart, N. Langellier, R. Trubko, D. R. Glenn, R. R. Fu and R. L. Walsworth, "Principles and techniques of the quantum diamond microscope," *Nanophotonics,* vol. 8, p. 209, 2019.

[13] I. Fescenko, A. Jarmola, I. Savukov, P. Kehayias, J. Smits, J. Damron, N. Ristoff, N. Mosavian and V. M. Acosta, "Diamond magnetometer enhanced by ferrite flux concentrators," *Phys. Rev. Research,* vol. 2, no. 2, p. 023394, 2020.

[14] B. Zhao, H. Guo, R. Zhao, F. Du, Z. Li, L. Wang, D. Wu, Y. Chen, J. Tang and J. Liu, "High-sensitivity three-axis vector magnetometry using the electron spin ensembles in single diamond," *IEEE Magnetics Letters,* pp. 1-1, 2019.

[15] J. M. Schloss, J. F. Barry, M. J. Turner and R. L. Walsworth, "Simultaneous Broadband Vector Magnetometry Using Solid-State Spins," *Physical Review Applied,* vol. 10, no. 3, p. 034044, 2018.

[16] M. Dale and G. Morley, "Medical applications of diamond magnetometry: commercial viability," *ArXiv e-prints,* #may# 2017.

[17] F. M. Stürner, A. Brenneis, J. Kassel, U. Wostradowski, R. Rölver, T. Fuchs, K. Nakamura, H. Sumiya, S. Onoda, J. Isoya and F. Jelezko, "Compact integrated magnetometer based on nitrogen-vacancy centres in diamond," *Diamond and Related Materials,* vol. 93, p. 59, 2019.

[18] J. L. Webb, J. D. Clement, L. Troise, S. Ahmadi, G. J. Johansen, A. Huck and U. L. Andersen, "Nanotesla sensitivity magnetic field sensing using a compact diamond nitrogen-vacancy magnetometer," *Applied Physics Letters,* vol. 114, p. 231103, 2019.

[19] D. Zheng, Z. Ma, W. Guo, L. Niu, J. Wang, X. Chai, Y. Li, Y. Sugawara, C. Yu, Y. Shi, X. Zhang, J. Tang, H. Guo and J. Liu, "A hand-held magnetometer based on an ensemble of nitrogen-vacancy centers in diamond," *Journal of Physics D: Applied Physics,* vol. 53, p. 155004, 2020.

[20] R. Patel, L. Zhou, A. Frangeskou, G. Stimpson, B. Breeze, A. Nikitin, M. Dale, E. Nichols, W. Thornley, B. Green, M. Newton, A. Edmonds, M. Markham, D. Twitchen and G. Morley, "Sub-nanotesla magnetometry with a fibre-coupled diamond sensor," *arXiv e-prints,* p. arXiv:2002.08255, 2020.

[21] D. Kim, M. I. Ibrahim, C. Foy, M. E. Trusheim, R. Han and D. R. Englund, "A CMOS-integrated quantum sensor based on nitrogen--vacancy centres," *Nature Electronics,* vol. 2, pp. 284-289, 2019.

[22] S. Maayani, C. Foy, D. Englund and Y. Fink, "Distributed Quantum Fiber Magnetometry," *Laser \& Photonics Reviews,* vol. 0, p. 1900075, 2019.

[23] V. M. Acosta, E. Bauch, M. P. Ledbetter, A. Waxman, L.-S. Bouchard and D. Budker, "Temperature Dependence of the Nitrogen-Vacancy Magnetic Resonance in Diamond," *Physical Review Letters,* vol. 104, no. 7, p. 070801, 2010.

[24] T. Plakhotnik, "Diamonds for quantum nano sensing," *Current Opinion in Solid State and Materials Science,* vol. 21, pp. 25-34, 2017.

[25] A. Gruber, A. Dräbenstedt, C. Tietz, L. Fleury, J. Wrachtrup and C. v. Borczyskowski, "Scanning Confocal Optical Microscopy and Magnetic Resonance on Single Defect Centers," *Science,* vol. 276, pp. 2012-2014, 1997.

[26] K. Bayat, J. Choy, M. Farrokh Baroughi, S. Meesala and M. Lončar, "Efficient, Uniform, and Large Area Microwave Magnetic Coupling to NV Centers in Diamond Using Double Split-Ring Resonators," *Nano Letters,* vol. 14, pp. 1208-1213, 2014.

[27] Z. Ma, S. Zhang, Y. Fu, H. Yuan, Y. Shi, J. Gao, L. Qin, J. Tang, J. Liu and Y. Li, "Magnetometry for precision measurement using frequency-modulation microwave combined efficient photon-collection technique on an ensemble of nitrogen-vacancy centers in diamond," *Optics Express,* vol. 26, pp. 382-390, 2018.

[28] S. Ahmadi, H. A. R. El-Ella, J. O. B. Hansen, A. Huck and U. L. Andersen, "Pump-Enhanced Continuous-Wave Magnetometry Using Nitrogen-Vacancy Ensembles," *Physical Review Applied,* vol. 8, no. 3, p. 034001, 2017.

[29] D. Duan, V. K. Kavatamane, S. R. Arumugam, G. Rahane, Y.-K. Tzeng, H.-C. Chang, H. Sumiya, S. Onoda, J. Isoya and G. Balasubramanian, "Enhancing fluorescence excitation and collection from the nitrogen-vacancy center in diamond through a micro-concave mirror," *Applied Physics Letters,* vol. 113, p. 041107, 2018.

[30] I. Fedotov, L. Doronina-Amitonova, A. Voronin, A. Levchenko, S. Zibrov, D. Sidorov-Biryukov, A. Fedotov, V. Velichansky and A. Zheltikov, "Electron spin manipulation and readout through an optical fiber," *Scientific Reports,* vol. 4, p. 5362, 2014.

[31] A. K. Dmitriev and A. K. Vershovskii, "Concept of a microscale vector magnetic field sensor based on nitrogen-vacancy centers in diamond," *Journal of the Optical Society of America B,* vol. 33, pp. B1--B4, 2016.

[32] L. Mayer and T. Debuisschert, "Direct optical interfacing of CVD diamond for deported sensing experiments involving nitrogen-vacancy centres," *physica status solidi (a),* vol. 213, pp. 2608-2613, 2016.

[33] H. A. R. El-Ella, S. Ahmadi, A. M. Wojciechowski, A. Huck and U. L. Andersen, "Optimised frequency modulation for continuous-wave optical magnetic resonance sensing using nitrogen-vacancy ensembles," *Optics Express,* vol. 25, pp. 14809-14821, 2017.

[34] K. Jensen, V. Acosta, A. Jarmola and D. Budker, "Light narrowing of magnetic resonances in ensembles of nitrogen-vacancy centers in diamond," *Physical Review B,* vol. 87, p. 014115, 2013.

[35] D. W. Allan, "Should the classical variance be used as a basic measure in standards metrology?," *IEEE Transactions on Instrumentation and Measurement,* Vols. IM-36, pp. 646-654, June 1987.

[36] H. J. Mamin, M. H. Sherwood, M. Kim, C. T. Rettner, K. Ohno, D. D. Awschalom and D. Rugar, "Multipulse Double-Quantum Magnetometry with Near-Surface Nitrogen-Vacancy Centers," *Physical Review Letters,* vol. 113, no. 3, p. 030803, 2014.

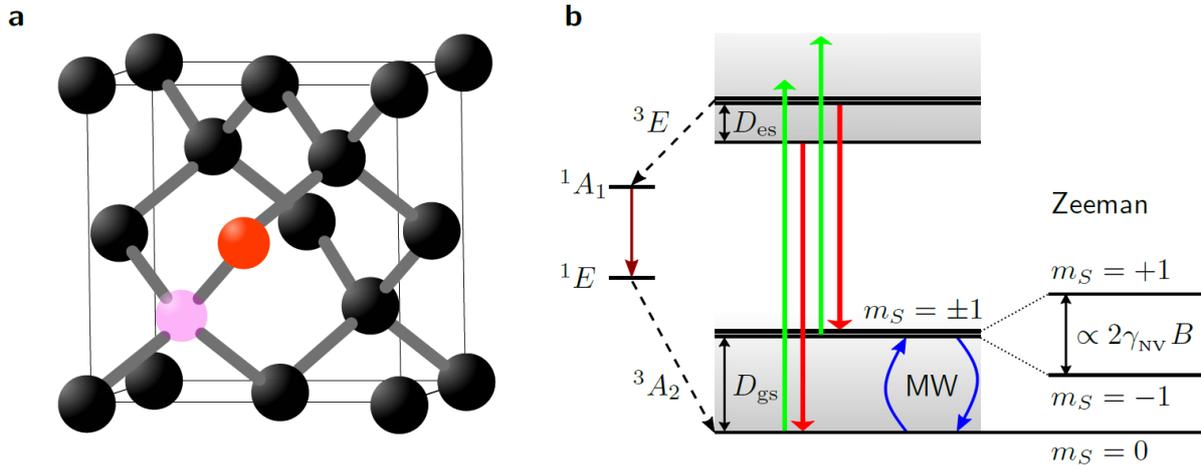

**Figure 1 NV center in diamond and its energy diagram.** a) The negatively charged nitrogen-vacancy center is a color center in the lattice of the diamond consisting of a nitrogen atom (red) and an adjacent vacancy (light purple). Carbon atoms are shown as black spheres. The NV center can be orientated along four possible crystallographic axes of the diamond. b) Schematic of the energy level scheme of the negatively charged NV center. Both the ground state ($^3A_2$) and the excited state ($^3E$) are spin triplets. The zero-field splitting energy of the ground state is $D_{gs} \approx 2.87$ GHz. The electron spin states of the $S = 1$ system are denoted with the spin quantum numbers $m_S = 0$ and $m_S = \pm 1$. Spin population can be optically pumped into the $^3E$ states by green light (green arrows), from which the spins decay back to $^3A_2$ under the emission of red fluorescence (red arrows), while the spin quantum number is conserved. Spins in the $m_S = \pm 1$ have a higher probability than those with $m_S = 0$ to undergo intersystem crossing via the states $^1A_1$ and $^1E$ with infrared emission (dark red arrow). This results in a reduced fluorescence emission and a polarization into the $m_S = 0$ ground state. A resonant microwave field enables population transfer between $m_S = 0$ and $m_S = \pm 1$ states. The $m_S = \pm 1$ states split under the application of an external magnetic field, which can be optically detected in an ODMR measurement.

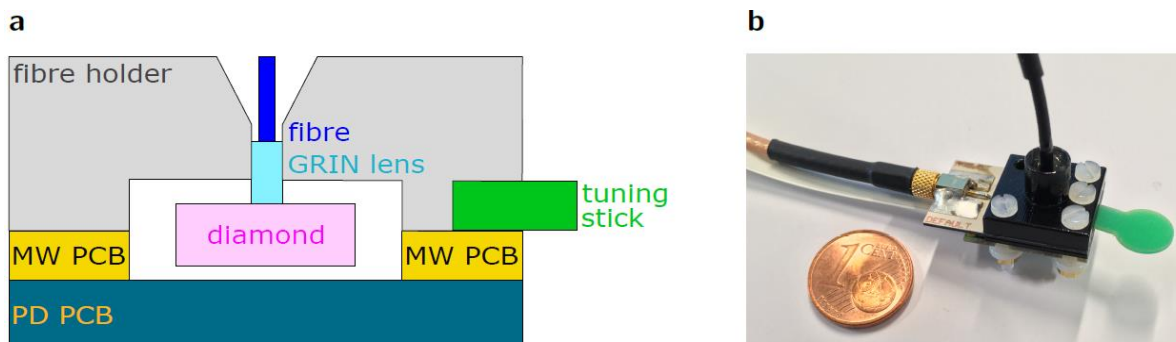

**Figure 2 Assembly of the sensor head.** a) Schematic cross section of the sensor head including the used components (optical fiber, GRIN rod lens, 3D-printed fiber holding construction, diamond, microwave board (MW PCB), resonator tuning stick, MMCX

connector, and the balanced photodetection board (PD PCB) including two photodiodes with different coating and electronic components at the back of this PCB. b) Photo image of the NV sensor placed next to a Euro cent coin.

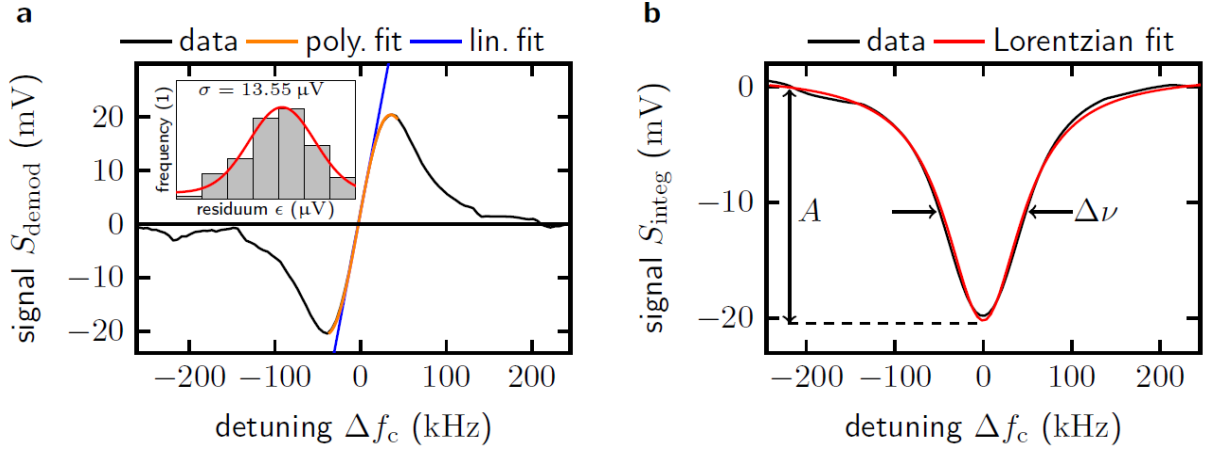

**Figure 3 Demodulated and integrated signal of an ODMR spectrum.** a) The recorded signal $S_\text{demod}$ (black curve) is shown as a function of the detuning $\Delta f_c$ from the resonance frequency, which is characterized by the zero-crossing of the signal. A polynomial of 6$^\text{th}$ order (orange curve) was applied as empirical fit to the data to determine the slope parameter at the zero-crossing. The blue curve shows the linear contribution of the polynomial. The noise parameter was derived by the standard deviation $\sigma$ of the residuum $\epsilon$ that results from subtracting the polynomial fit values from the data, which are summarized in a histogram. b) Integrating $S_\text{demod}$ resembles the resonance curve $S_\text{integ}$ (black curve) with a Lorentzian shape (fit red curve) with $\Delta\nu$ the linewidth and $A$ the amplitude of the resonance linked to contrast $C$.

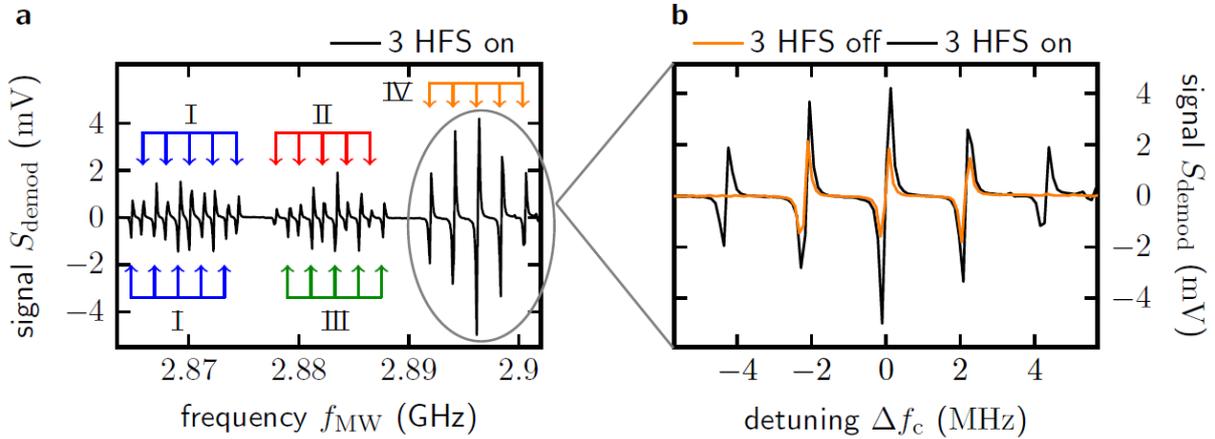

**Figure 4 Optically detected magnetic resonance (ODMR).** a) Spectrum of the measured demodulated signal $S_\text{demod}$ showing spin transitions from the $m_s = 0$ to $m_s = 1$ states including hyperfine interactions. The spectrum was recorded at a modulation frequency of 5 kHz and a modulation depth of 100 kHz with 200 ms integration time. The total microwave power was 0.91 mW with a simultaneous excitation of the three hyperfine split transitions. Each NV axis produces five characteristic zero-crossings resulting from the simultaneous manipulation of the $^{14}$N hyperfine splitting, which are labelled with colored arrows and Roman letters. An exception is orientation I, where the transitions from $m_s = 0$ to $m_s = \pm 1$ overlap around $D_\text{gs} \approx 2.87$ GHz, so that ten characteristic crossings are visible. b) Detailed view of the measured resonances of the NV orientation with the largest magnetic field

projection with a comparison of the applied microwave manipulation: '3 HFS on' (black curve) and '3 HFS off' (orange curve).

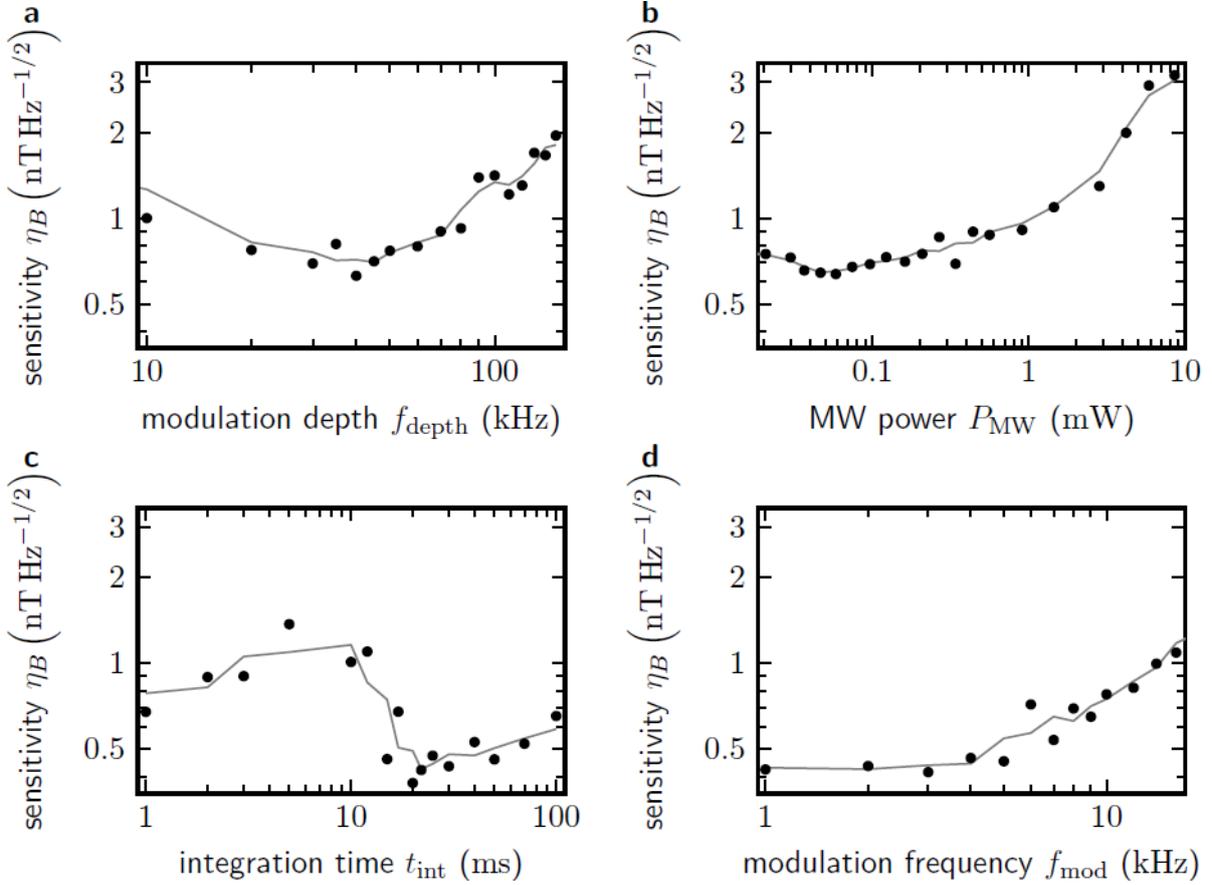

**Figure 5 Evaluation of the magnetic field sensitivity with respect to sensor parameters.** The sensitivity was evaluated on the NV orientation with the largest field projection, *i.e.* NV orientation IV in Figure 4a. The gray line corresponds to a moving mean of three neighboring data points and is added to the graphs as a guide to the eye. a) Variation of the modulation depth $f_{\text{depth}}$ at a microwave power of 16 µW, 1 kHz modulation frequency for an integration time of 1 ms. b) Variation of the microwave power $P_{\text{MW}}$ at 1 kHz modulation frequency and 40 kHz modulation depth for an integration time of 1 ms. c) Variation of the integration time $t_{\text{int}}$ at a microwave power of 58 µW, 1 kHz modulation frequency, and 40 kHz modulation depth. d) Variation of the modulation frequency $f_{\text{mod}}$ at a microwave power of 58 µW and 40 kHz modulation depth for an integration time of 20 ms.

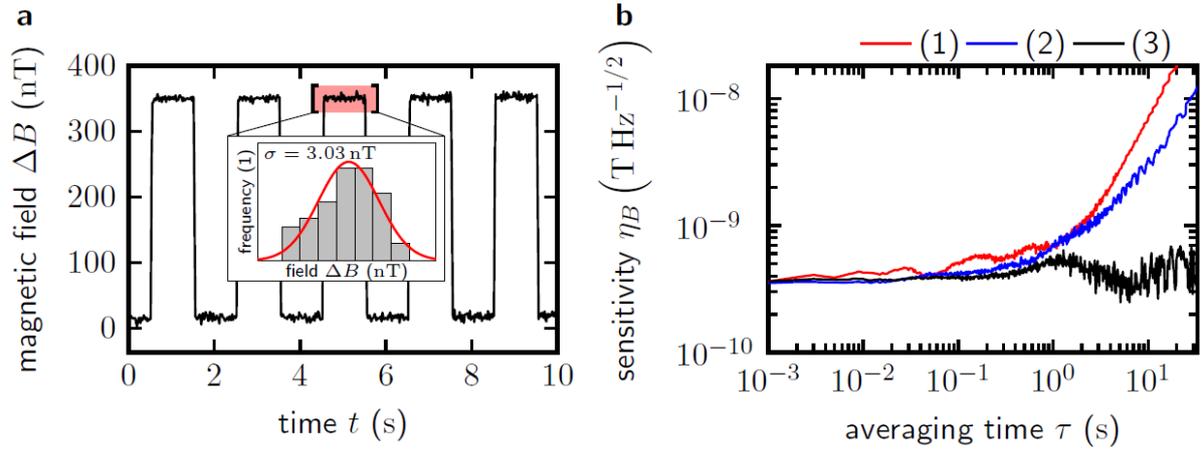

**Figure 6 Sensitivity determination with square-wave test field using the Allan deviation.**
a) The demodulated signal $S_{\text{demod}}$ was recorded for 10 s with an integration time of 20 ms per data point. A test magnetic field was applied with a square-wave function of 2 Hz and amplitude of 2.18 mA on the copper windings around the pair of Helmholtz coils corresponding to an induced magnetic field of 333 nT and is clearly visible in the recorded data trace (black curve). The noise on a single 1 s step was evaluated by the standard deviation of a Gaussian distribution ($\sigma = 3.03$ nT, red curve) on the recorded data and extrapolates to 429 pT Hz$^{-1/2}$ for 20 ms integration time. b) Magnetic field sensitivity derived by analyzing the Allan deviation $\sigma_A(\tau)$ of recorded time traces (without applying any magnetic test field). The data of (1) and (2) are based on time traces of magnetically sensitive cases, where the applied microwave field was set to the point of zero-crossing in the demodulated spectrum. In (2) the NV sensor was totally enclosed in the magnetic shielding of two half-shells made of mu-metal, for details see main text, whereas in (1) the sensor was unshielded. (3) shows the magnetically insensitive data without using any shielding. The time traces were recorded with a sampling rate of 1 kHz and a simultaneous manipulation of the three hyperfine states was applied in all measurements.

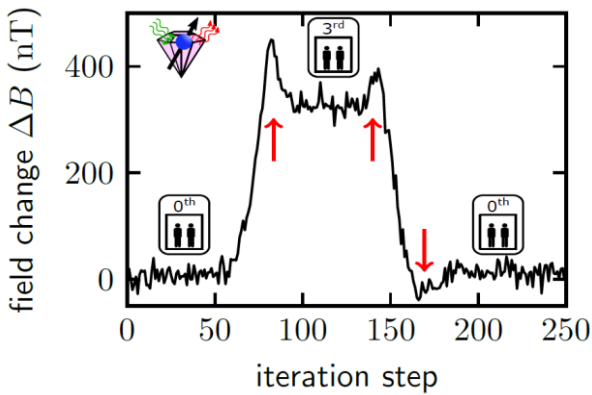

**Figure 7 Measurement of the change in the magnetic field due to the movement of an elevator.** The NV magnetometer was positioned roughly 4 m away from the doors of an elevator three floors above the ground floor. For tracking the resonance condition of the selected NV transition, the demodulated signal $S_{\text{demod}}$ was used as an error signal for a feedback loop to adjust the applied microwave frequency. The microwave frequency was adjusted by 250 Hz corresponding to a magnetic field change of 8.92 nT. We recorded $S_{\text{demod}}$ for 1 ms integration time (per iteration step) with the following experimental parameters: 1 kHz modulation frequency, 40 kHz modulation depth, and 58 µW microwave power. In this measurement, the elevator started in the ground floor and moved to third level, opened and closed its doors, and moved then down to the ground floor. The red arrows indicate the impact of opening and closing the elevator's doors.

**Table of contents entry**
Negatively charged nitrogen-vacancy centers in diamond are a promising candidate for high sensitive magnetometers enabling wide range of applications. Transferring this technology into products, this sensor needs to be miniaturized and integrated such that it can be portably used without the requirement of bulky laboratory equipment. A portable prototype of a fiber-integrated magnetometer with subnanotesla sensitivity is demonstrated.

Keywords
Quantum technologies, quantum sensing, magnetometers, diamonds, nitrogen vacancy centers

Authors
Felix M. Stürner, Andreas Brenneis*, Thomas Buck, Julian Kassel, Robert Rölver, Tino Fuchs, Anton Savitsky, Dieter Suter, Jens Grimmel, Stefan Hengesbach, Michael Förtsch, Kazuo Nakamura, Hitoshi Sumiya, Shinobu Onoda, Junichi Isoya, Fedor Jelezko

Title
Integrated and portable magnetometer based on nitrogen-vacancy ensembles in diamond

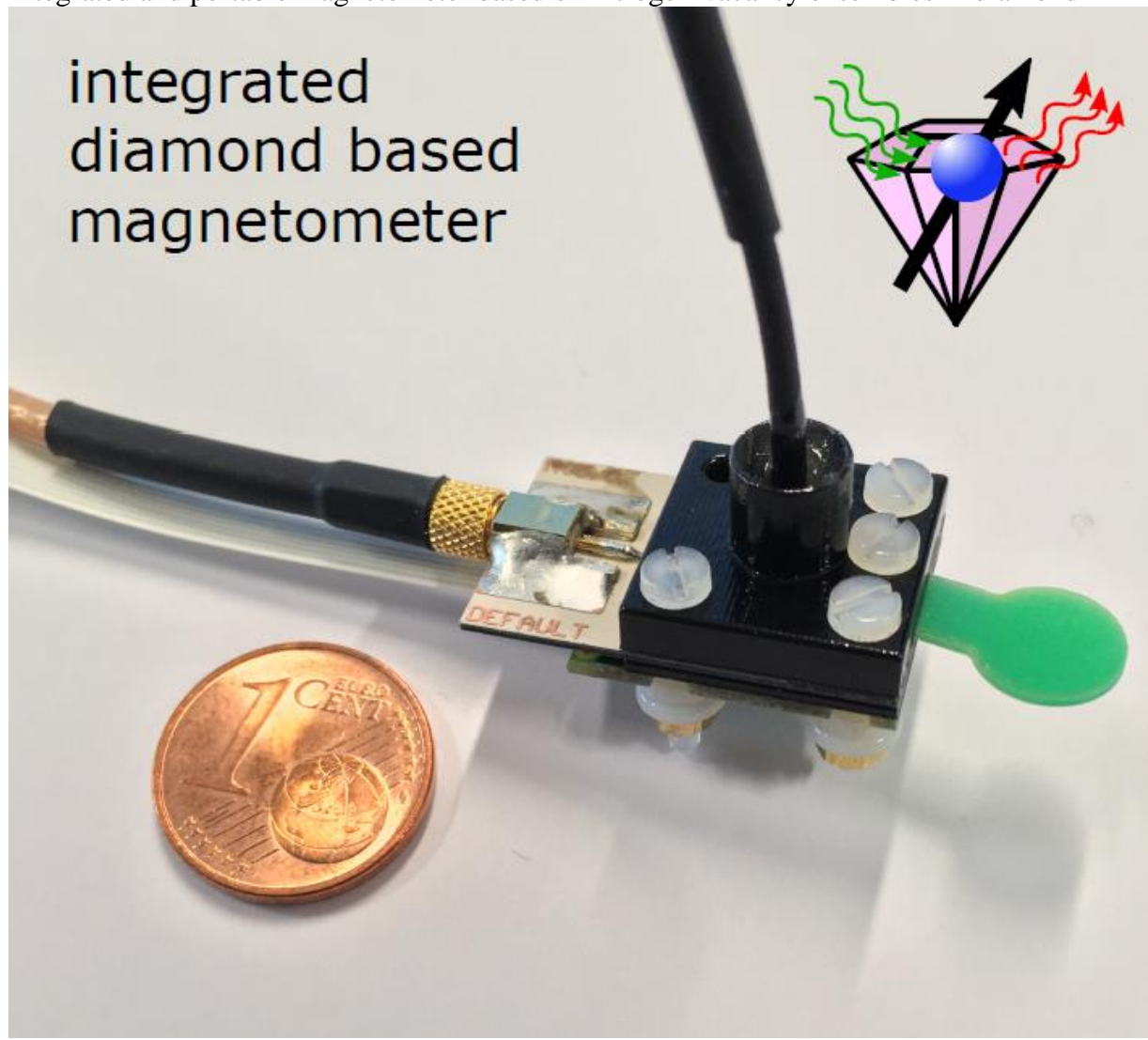

# Supplementary Information:
# Integrated and portable magnetometer based on nitrogen-vacancy ensembles in diamond

*Felix M. Stürner, Andreas Brenneis,* Julian Kassel, Thomas Buck, Robert Rölver, Tino Fuchs, Anton Savitsky, Dieter Suter, Jens Grimmel, Stefan Hengesbach, Michael Förtsch, Kazuo Nakamura, Hitoshi Sumiya, Shinobu Onoda, Junichi Isoya, Fedor Jelezko*

## S1   Setup of the NV magnetometer

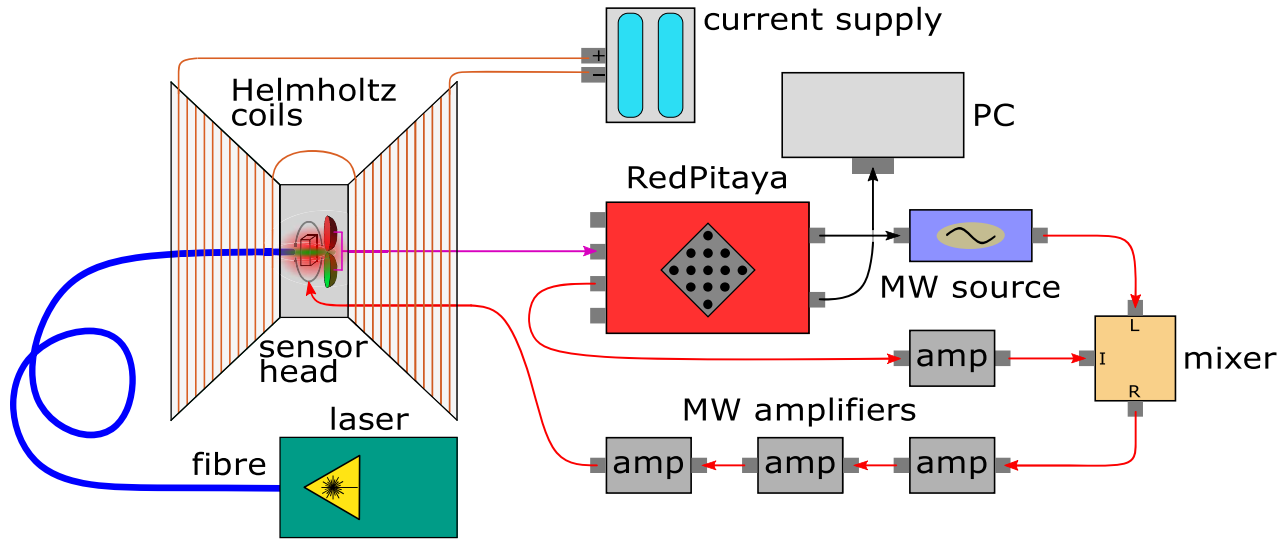

**Figure S1: Schematic of the complete NV sensor setup.** The sensor head of our NV magnetometer was placed inside a tapered pair of Helmholtz coils, which were powered by a battery. The excitation light was generated by a compact laser module and transmitted to the diamond by a single mode fiber. A local oscillator generated microwaves in the GHz range and a mixer was used to add microwaves in the MHz range, which were generated and frequency modulated by the waveform generator module of the RedPitaya. These microwaves were further amplified by three compact low-noise amplifiers and then transmitted to the double split-ring resonator providing an oscillating field at the diamond necessary for the driving electron spin resonances. A balanced detection scheme consisting of two photodiodes and a logarithmic transimpedance amplifier was used to record both the modulated fluorescence signal and laser light. The output of the transimpedance amplifier was demodulated by the RedPitaya module and further processed by a personal computer.

# S2 Optical Excitation

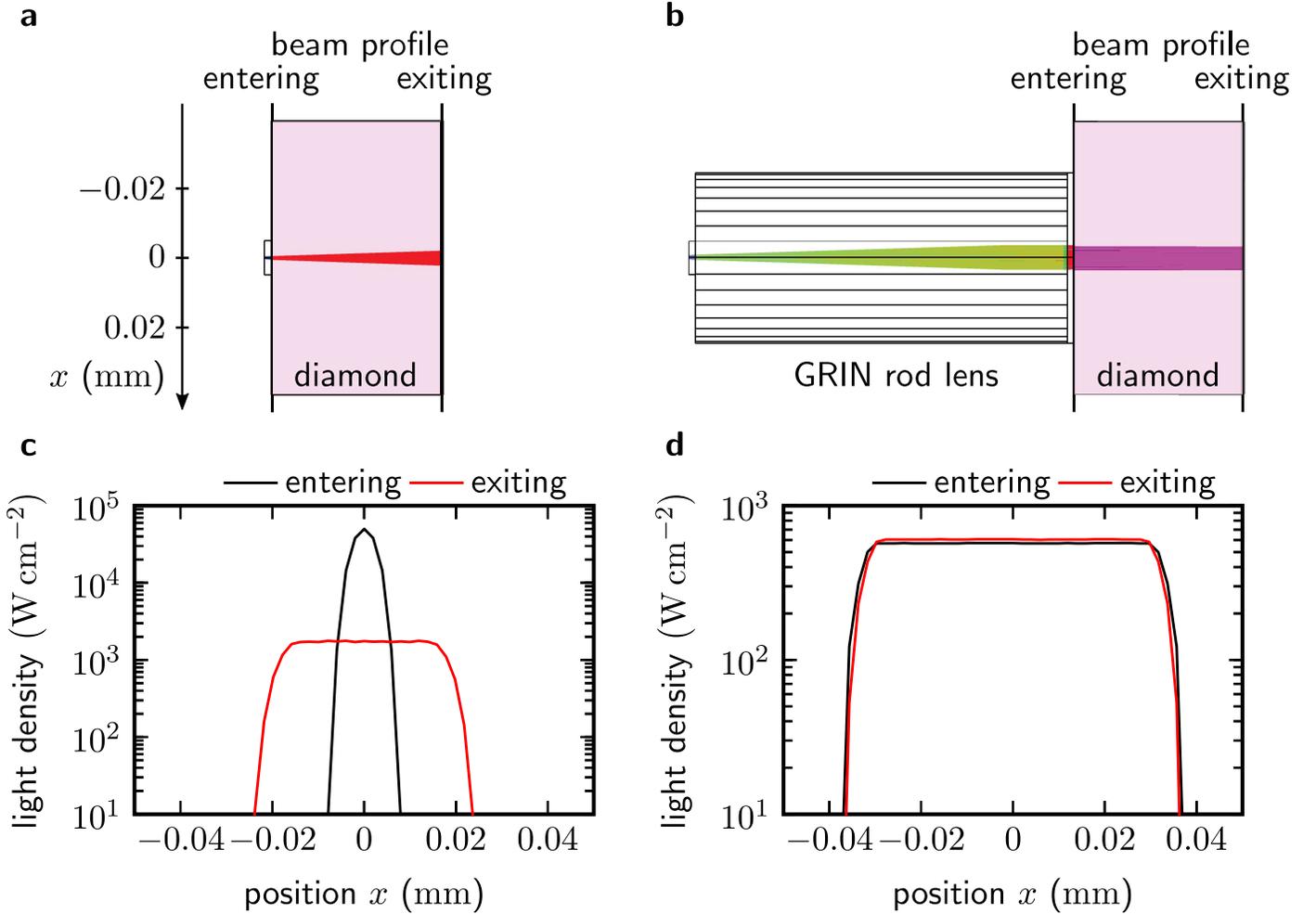

Figure S2: **Optical ray tracing simulations for the optical excitation of the NV centers.** Optical ray trace simulations were performed using Zemax OpticStudio®. We used a diamond with dimensions of $0.8 \times 0.8 \times 0.5\,\mathrm{mm}^3$ and modeled the optical output of a single mode fiber with numerical aperture $NA_\mathrm{fiber} = 0.12$ by a radial emitting area of $2.0 \times 2.0\,\mathrm{\mu m}^2$. The emitting surface was embedded in a block of polydimethylsiloxane (PDMS) with refractive index of $n_\mathrm{PDMS} = 1.46$ and emitted light rays with an optical power $P_\mathrm{out} = 23.5\,\mathrm{W}$ within an angle $\theta$ of $4.813°$ resulting from $NA_\mathrm{fiber} = n_\mathrm{PDMS}\sin(\theta)$. We evaluated the beam profile at two positions inside the diamond: $1\,\mathrm{\mu m}$ deep from the surface of the diamond, which is connected to the fiber-GRIN compound (entering beam profile) and $499\,\mathrm{\mu m}$ deep along the optical axis (exiting beam profile). a) shows the graphical 2D view on the simulation settings of a direct light coupling between fiber and diamond, whereas b) shows the 2D view, which corresponds to the sensor head of the NV magnetometer using a GRIN rod lens inserted in the optical path between fiber output and diamond with an additional connection layer of PDMS between lens and diamond. For reasons of clarity, the axis of the position $x$ describing the lateral distance to the center of the fiber and used to evaluate the light distribution inside the diamond is only shown in a). Simulation results of the optical illumination evaluated at the two positions (entering and exiting the diamond) inside the diamond without using a GRIN rod lens shown in c) and with using a GRIN in lens d). A more homogeneous illumination of the diamond is achieved for b) as the light rays leaving the fiber are parallelized by the GRIN rod lens.

## S3 Offset magnetic field

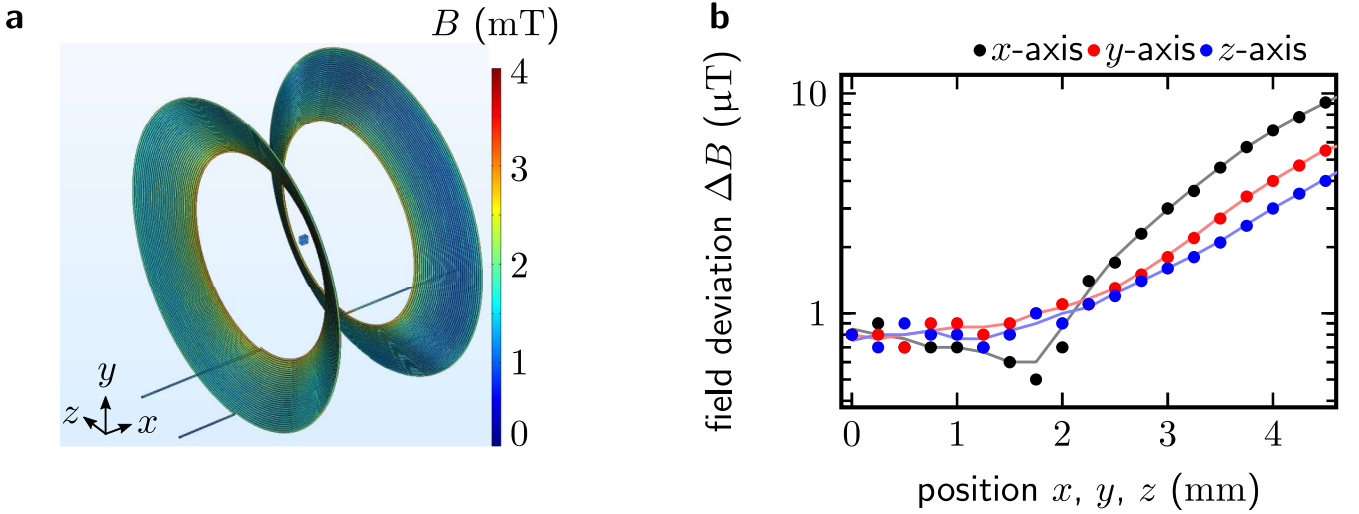

**Figure S3: Simulative evaluation of the offset magnetic field generated by a tapered Helmholtz coil.** a) Graphical view of the simulation settings for evaluating the generated magnetic field of the tapered Helmholtz coil using COMSOL Multiphysics®. The field distribution was evaluated inside the volume of the diamond (with dimensions of $0.8 \times 0.8 \times 0.5\,\text{mm}^3$). The diamond was positioned at the center of the two coils with its largest surface in the $xz$-plane. A current of $I = 500\,\text{mA}$ was applied to the coils, which were electrically connected with each other. In this graphic, the generated magnetic field strength $B$ at the surface of the coils is plotted. b) The position of the diamond relative to the center of the two coils was varied along the $x$-, $y$-, and $z$-axis and the simulated field deviation of $B$ inside the volume of the diamond is shown, respectively. Lines of moving mean (of three neighboring data points) are added to the graphs.

## S4 Signal recording and signal processing

As described in the main text, we used a balanced detection scheme consisting of two photodiodes and a logarithmic transimpedance amplifier to record both the modulated fluorescence signal and laser light. The output voltage $V_{\text{out}}$ of the transimpedance amplifier is given by

$$V_{\text{out}} = G \cdot K \cdot \log_{10}\left(\frac{I_{\text{sig}} + \Delta I_{\text{sig}}}{I_{\text{ref}}}\right), \tag{S1}$$

where $G$ is a gain factor with $G = 20.66\,\text{A}^{-1}$, $K$ the logarithmic scale factor with $K = 0.375\,\text{V}$, $I_{\text{sig}}$ the measured photocurrent that corresponds to the background of the fluorescence, $\Delta I_{\text{sig}}$ the change of the photocurrent that corresponds to the drop of the fluorescence under resonant excitation and is caused by the modulation of the microwave frequency. $I_{\text{ref}}$ is the measured photocurrent of the reference photodiode. Using a Taylor expansion the output voltage $V_{\text{out}}$ can be approximated to

$$V_{\text{out}} \approx G \cdot K \cdot \left[\log_{10}\left(\frac{I_{\text{sig}}}{I_{\text{ref}}}\right) + \frac{\Delta I_{\text{sig}}}{I_{\text{sig}}} \cdot \frac{1}{\ln(10)} - \frac{1}{2} \cdot \left(\frac{\Delta I_{\text{sig}}}{I_{\text{sig}}}\right)^2 \cdot \frac{1}{\ln(10)}\right] \tag{S2}$$

$$\approx G \cdot K \cdot \log_{10}\left(\frac{\Delta I_{\text{sig}}}{I_{\text{ref}}}\right) + G \cdot K \cdot \frac{\Delta I_{\text{sig}}}{I_{\text{sig}}} \cdot \frac{1}{\ln(10)} + \mathcal{O}(n^2). \tag{S3}$$

From this equation, the modulated fluorescence signal of the NV centers can be extracted and corresponds to the voltage difference $\Delta V_{\text{out}}$ with

$$\Delta V_{\text{out}} \approx G \cdot K \cdot \frac{\Delta I_{\text{sig}}}{I_{\text{sig}}} \cdot \frac{1}{\ln(10)}. \tag{S4}$$

In order to obtain the electron spin resonance of a standard ODMR measurement, we integrated the demodulated signal $S_{\text{demod,i}}$ of the $i$ applied microwave frequencies (with $i \in \mathcal{N}$ and step size of $\Delta f_{\text{step}}$) over the measured spectral range by

$$S_{\text{integ}} = \sum_i \frac{S_{\text{demod},i}}{2f_{\text{depth}}} \cdot \Delta f_{\text{step}}. \tag{S5}$$

Here, the division factor of two times the modulation depth $f_{\text{depth}}$ takes the applied frequency modulation into account. We extracted information about the linewidth $\Delta\nu$ and contrast $C$ of the electron spin resonance by fitting the resulting data $S_{\text{integ}}$ by a Lorentzian function $g(x)$ of

$$g(x) = A\frac{(\Delta\nu/2)^2}{(x-x_0)^2 + (\Delta\nu/2)^2}, \tag{S6}$$

where $A$ corresponds to $\Delta V_{\text{out}}$, $\Delta\nu$ the full width at half maximum of the resonance curve, and $x_0$ the resonance frequency (Figure 3b). With the knowledge of the photocurrent $I_{\text{sig}}$ corresponding to the background of the fluorescence, the contrast $C$ of the spin resonance can be expressed by the ratio of $\Delta I_{\text{sig}}$ and $I_{\text{sig}}$.

## S5 ODMR spectra

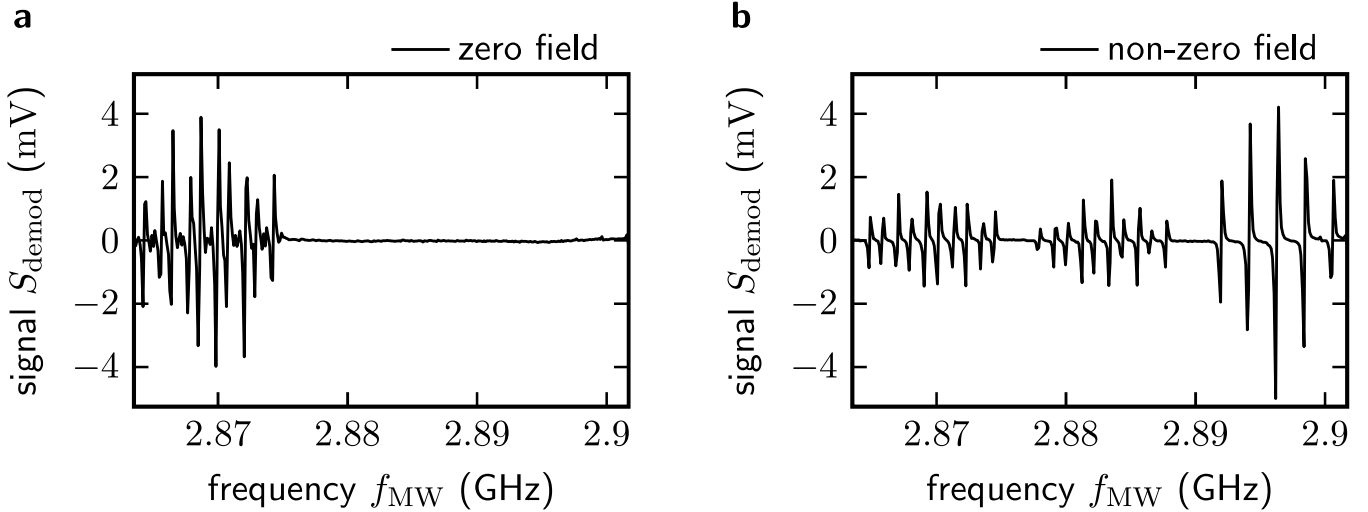

**Figure S4: Optically detected magnetic resonance (ODMR).** Signals were recorded with an integration time of 40 ms, a modulation depth $f_{\text{depth}}$ of 100 kHz, a modulation frequency $f_{\text{mod}}$ of 5 kHz, and a simultaneous manipulation of the three hyperfine splittings with a microwave power of 0.91 mW. The demodulated signal is shown in a) at zero field and in b) with an offset magnetic field generated by the tapered Helmholtz coils.

# S6 Spin manipulation

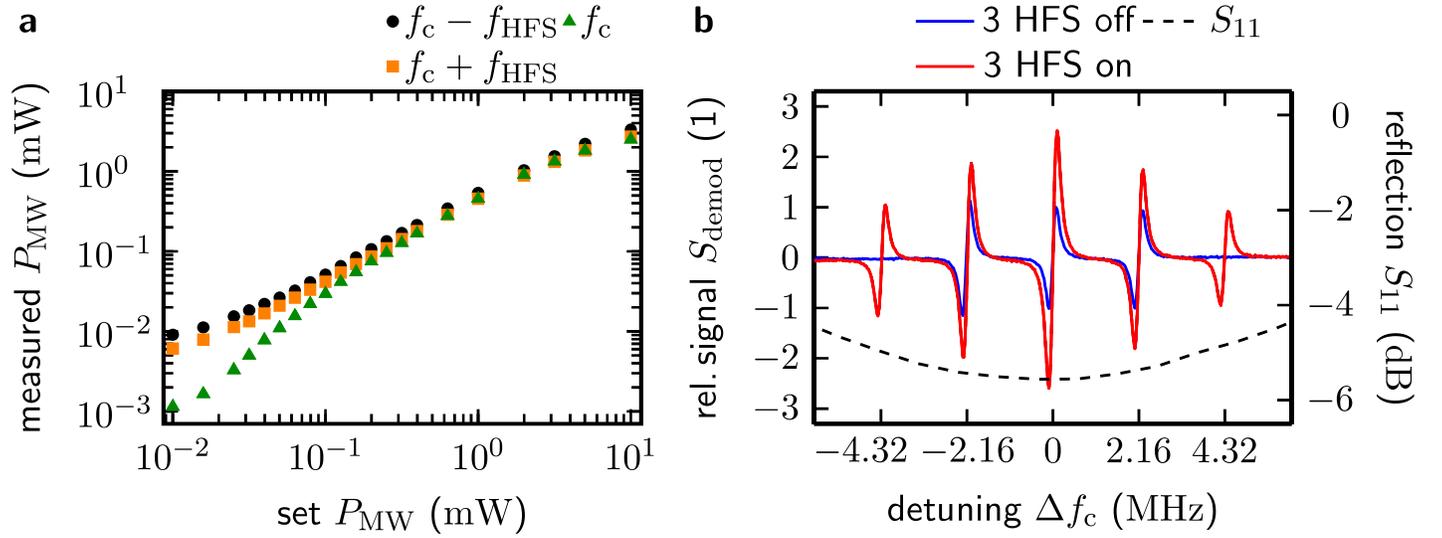

**Figure S5: Spin manipulation and ODMR spectrum for one NV orientation.** a) Power level of the applied frequencies for a spin manipulation with simultaneous spin manipulation measured by a spectrum analyzer (Rhode&Schwarz FSEA 30). The frequency of the local oscillator $f_{\mathrm{LO}}$ was set to 2.92 GHz, the baseband frequency $f_{\mathrm{BB}}$ to 30 MHz, and a hyperfine splitting frequency $f_{\mathrm{HFS}}$ of 2.16 MHz was used. The power levels were measured with a resolution bandwidth of 10 kHz. b) Comparison of the recorded ODMR spectra for a '3 HFS on' (red curve) and a '3 HFS off' (blue curve) spin manipulation. The spectra were recorded with a modulation frequency of 10 kHz and a modulation depth of 100 kHz with 1 ms integration time (data points were 200 times averaged). The microwave power was set to 0.91 mW. The measured signals $S_{\mathrm{demod}}$ were normalized to the peak amplitude of the central resonance around $f_{\mathrm{c}}$ of the '3 HFS off' measurement. Data of the reflection parameter $S_{11}$ of the double split-ring resonantor measured by a network analyzer (Hewlett Packard 8753D) is added to this plot (black dotted curve).

## S7 Logarithmic transimpedance amplification circuit

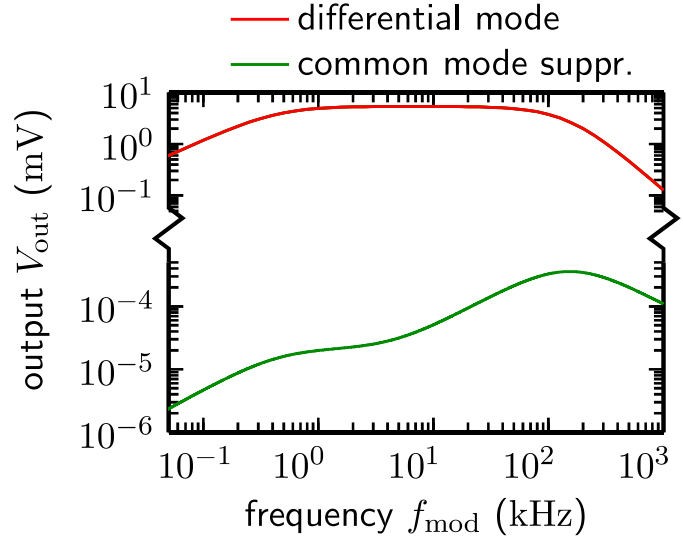

**Figure S6: Simulative analysis of the logarithmic transimpedance amplification circuit.** LTspice® was used to model the transimpedance amplification circuit including the utilized logarithmic transimpedance amplifier (Texas Instruments LOG114) and simulate the output voltage $V_{\text{out}}$ depending on the applied modulation frequency $f_{\text{mod}}$ and the input currents on the two photodiodes (signal photodiode $PD_{\text{sig}}$ used for detecting the fluorescence light and photodiode $PD_{\text{ref}}$ for referencing the laser light). In order to analyze the performance of the differential mode of this circuit, $V_{\text{out}}$ was evaluated for an input current $I_{\text{sig}}$ on $PD_{\text{sig}}$ with a sine-wave function (1 mA DC offset, 1 µA amplitude, and 100 kHz frequency) and a constant input current $I_{\text{ref}}$ (1 mA DC offset) on $PD_{\text{ref}}$ (red curve). The simulated 3 dB cut-off frequencies were at 430 Hz and at 103 kHz. For $f_{\text{mod}}$ in the range of 1 kHz–50 kHz a flat curve of $V_{\text{out}}$ with an average value of approximately 5.25 mV was obtained. The drop of the output voltage in the low frequency range around the lower cut-off frequency is mainly due to the implemented high-pass filter, whereas the drop of $V_{\text{out}}$ above the upper cut-off frequency is linked to the internal electronic characteristics of the transimpedance amplifier. The response $V_{\text{out}}$ was further evaluated for the common mode suppression by applying $I_{\text{ref}}$ on $PD_{\text{ref}}$ with a sine-wave function (1 mA DC offset, 1 µA amplitude, and 100 kHz frequency), whereas $I_{\text{sig}}$ was kept constant with 1 mA DC offset (green curve).

# S8 Analysis of the signal slope

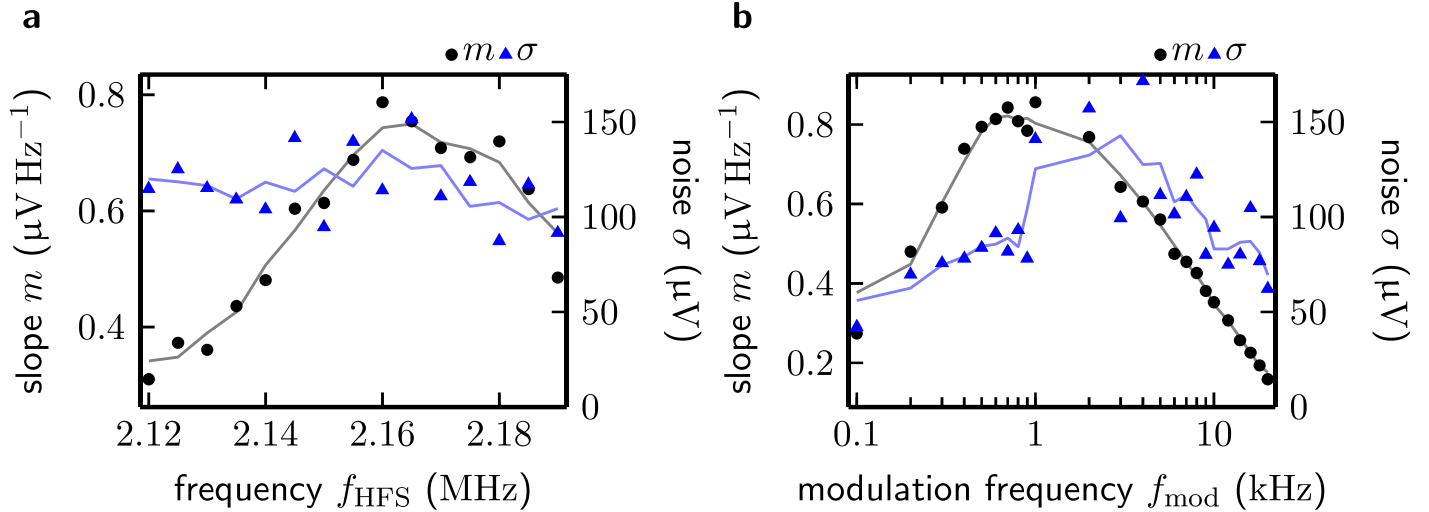

**Figure S7: Slope and noise dependence on the applied hyperfine splitting frequency $f_{\text{HFS}}$ and on the modulation frequency $f_{\text{mod}}$.** a) Slope $m$ (black dots) and noise $\sigma$ (blue triangles) parameters were extracted from a 6$^{\text{th}}$ order polynomial fit to the recorded signals $S_{\text{demod}}$ (see main text) for different hyperfine splitting frequencies $f_{\text{HFS}}$. These signals were recorded for an integration time of 20 ms, a modulation depth of 40 kHz, a modulation frequency of 1 kHz, and a simultaneous manipulation of the three hyperfine splittings with a microwave power of 58 µW. A line of moving mean (of three neighboring data points) is added to the graphs. b) Slope (black dots) and noise (blue triangles) values were analyzed based on the used modulation frequency. The data was recorded for an integration time of 20 ms, a modulation depth $f_{\text{depth}}$ of 40 kHz, and a simultaneous manipulation of the three hyperfine splittings with a microwave power of 1.45 mW. Lines of moving mean (of three neighboring data points) are added to the graphs.

## S9 Evaluation of the magnetic field sensitivity with respect to the sensor parameters

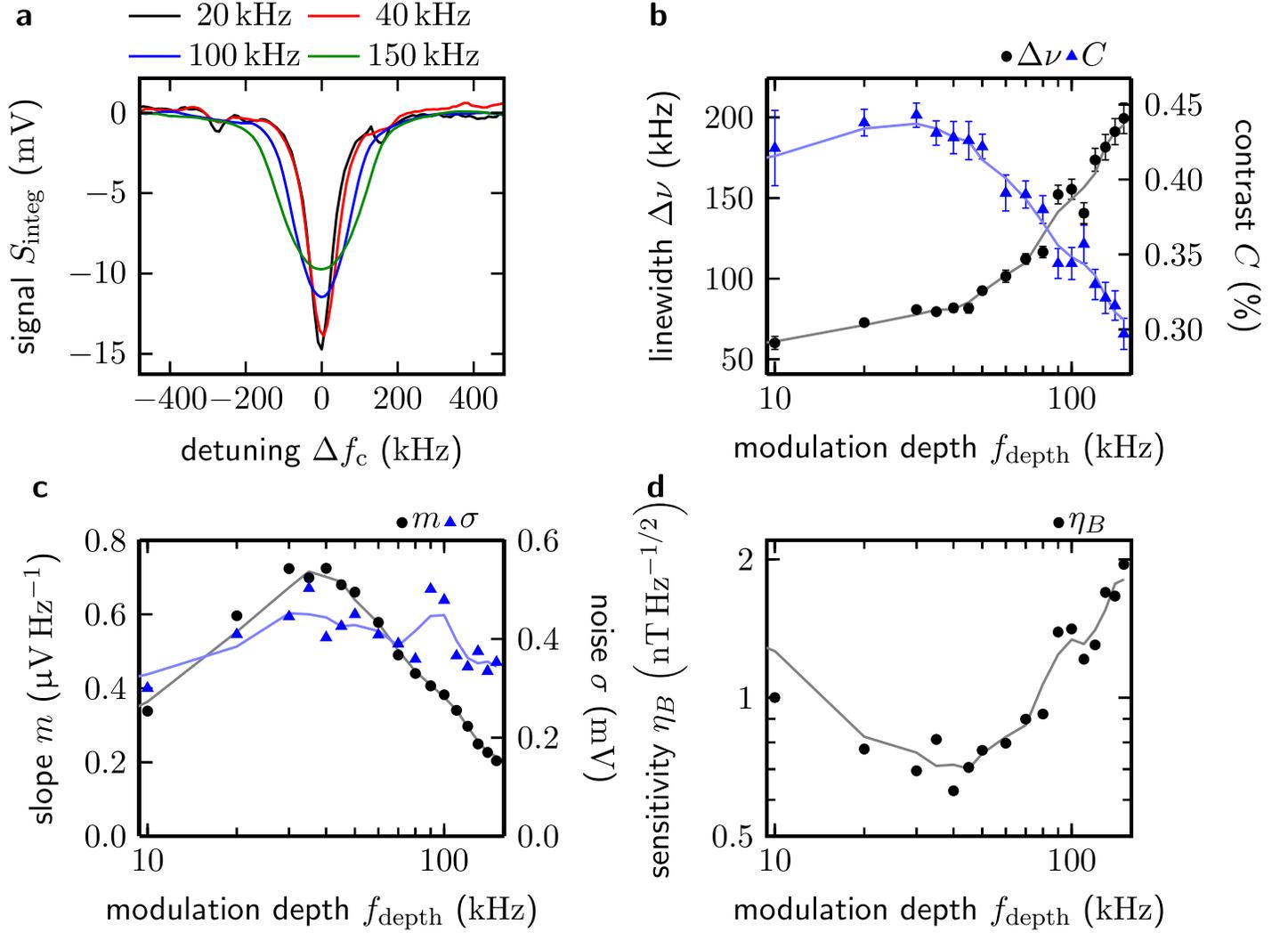

Figure S8: **Variation of the modulation depth to optimize the magnetic field sensitivity.** The data of this parameter variation was recorded with an integration time of 1 ms, a modulation frequency of 1 kHz, and a simultaneous manipulation of the three hyperfine splittings with a microwave power of 16 µW. Lines of moving mean (of three neighboring data points) are added to the graphs shown in b) to d). a) Plots of the integrated demodulated signal $S_{\text{integ}}$ for four applied modulation depths. b) Resulting linewidth $\Delta\nu$ (black dots) and contrast $C$ (blue triangles) depending on the modulation depth. c) Resulting slope $m$ (black dots) and noise $\sigma$ (blue triangles) depending on the modulation depth. d) Based on the measured slope and noise, the sensitivity $\eta_B$ (black dots) dependence on the modulation depth is shown.

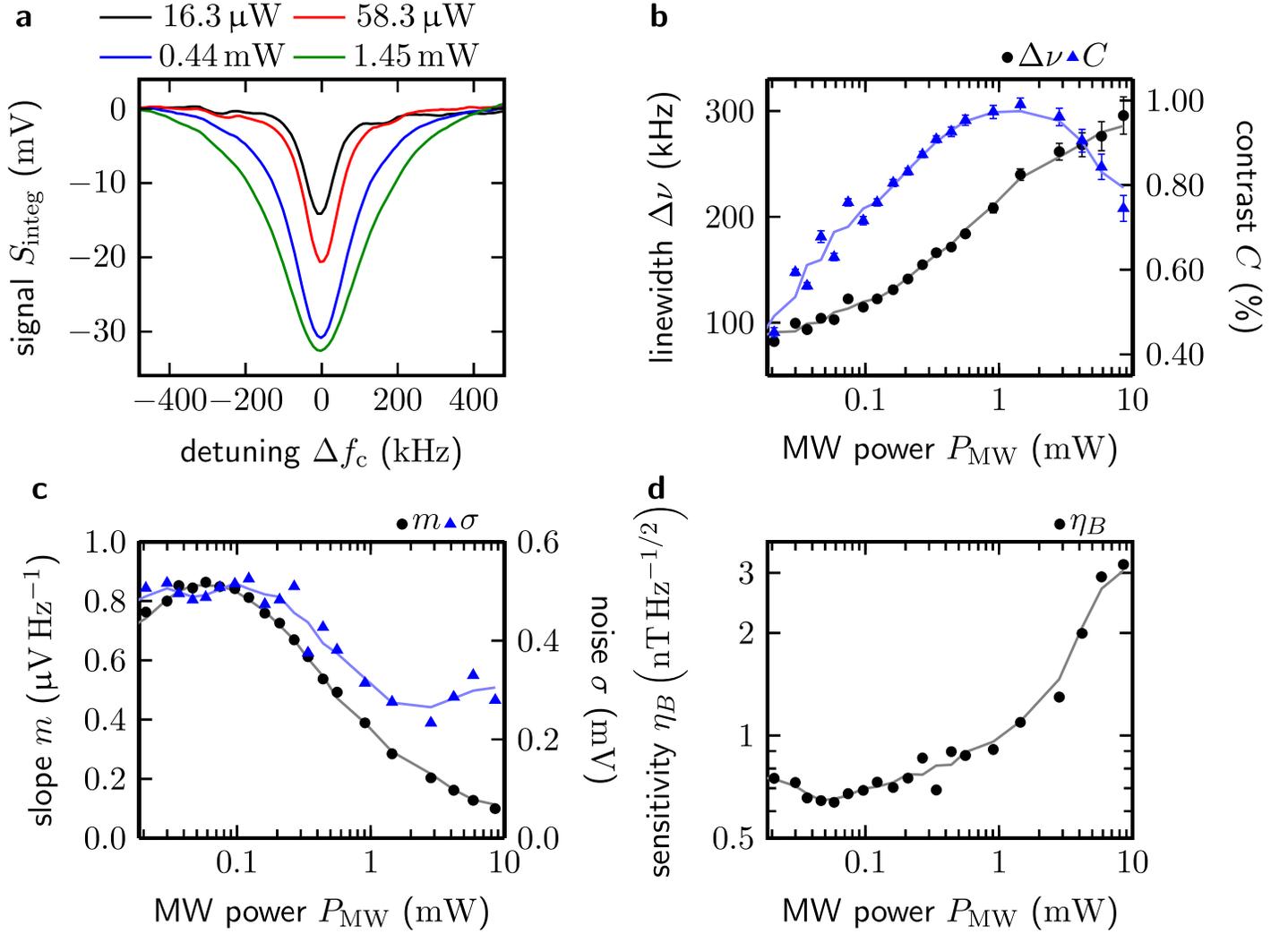

Figure S9: **Variation of the microwave power to optimize the magnetic field sensitivity.** The data of this parameter variation was recorded with an integration time of 1 ms, a modulation frequency of 1 kHz, 40 kHz modulation depth, and a simultaneous manipulation of the three hyperfine splittings. Lines of moving mean (of three neighboring data points) are added to the graphs shown in b) to d). a) Plots of the integrated demodulated signal $S_{\text{integ}}$ for four applied microwave powers. b) Resulting linewidth $\Delta\nu$ (black dots) and contrast $C$ (blue triangles) depending on the microwave power. c) Resulting slope $m$ (black dots) and noise $\sigma$ (blue triangles) depending on the microwave power. d) Based on the measured slope and noise, the sensitivity $\eta_B$ (black dots) dependence on the microwave power is shown.

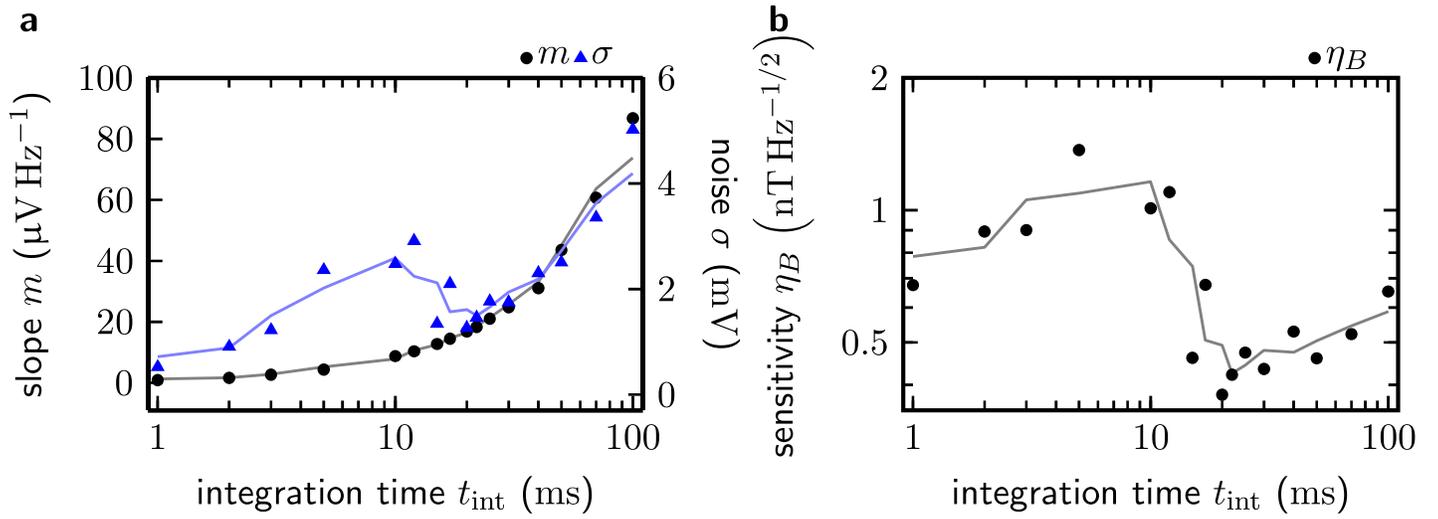

Figure S10: **Variation of the integration time to optimize the magnetic field sensitivity.** The data of this parameter variation was recorded with a modulation frequency of 1 kHz, a modulation depth of 40 kHz, and a simultaneous manipulation of the three hyperfine splittings with a microwave power of 58 μW. Lines of moving mean (of three neighboring data points) are added to the graphs shown in a) and b). a) Resulting slope $m$ (black dots) and noise $\sigma$ (blue triangles) depending on the integration time. b) Based on the measured slope and noise, the sensitivity $\eta_B$ (black dots) dependence on the integration time is shown.

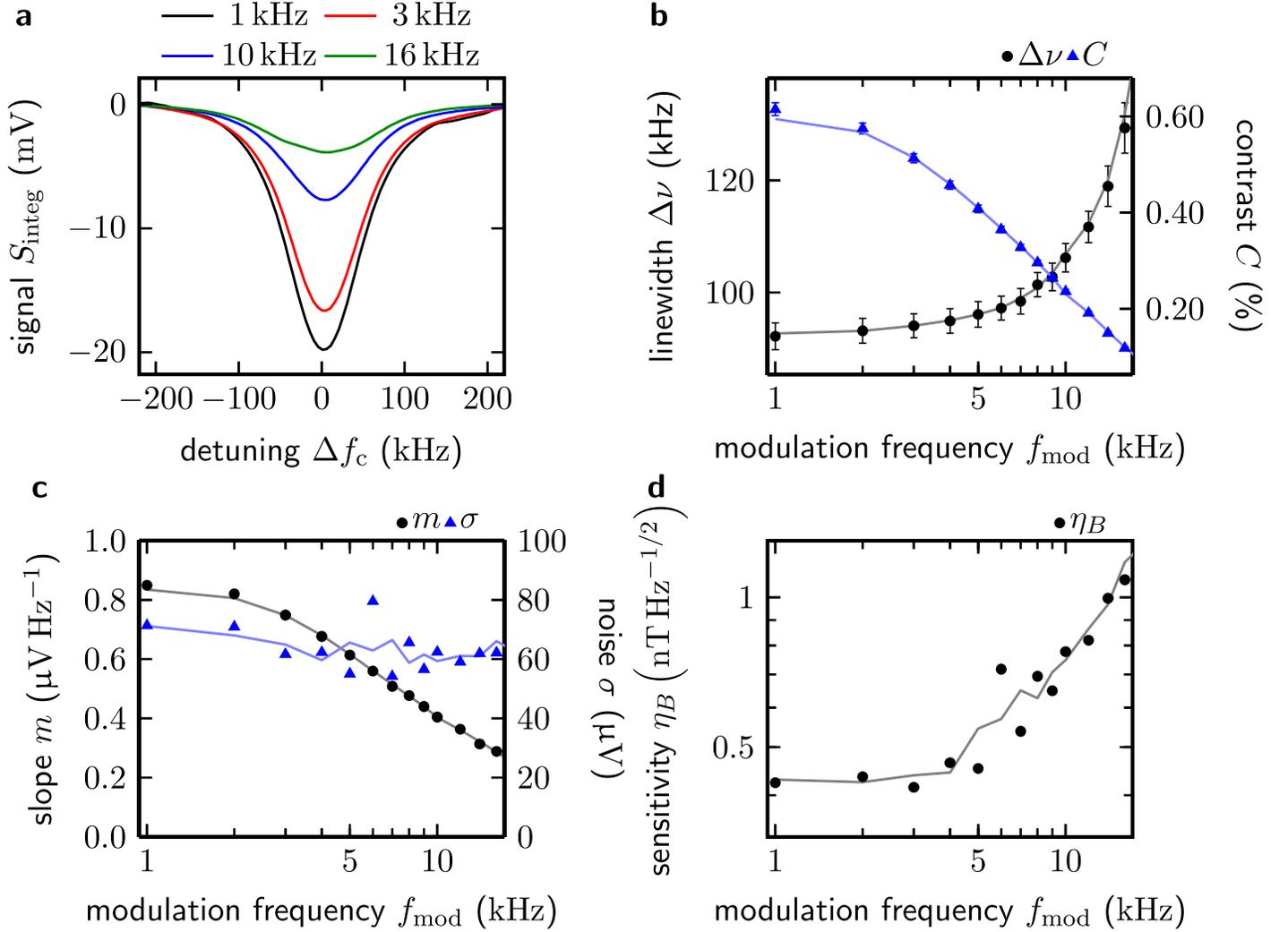

Figure S11: **Variation of the modulation frequency to optimize the magnetic field sensitivity.** The data of this parameter variation was recorded with an integration time of 20 ms, 40 kHz modulation depth, and a simultaneous manipulation of the three hyperfine splittings with a microwave power of 58 μW. Lines of moving mean (of three neighboring data points) are added to the graphs shown in b) to d). a) Plots of the integrated demodulated signal $S_{integ}$ for four applied modulation frequencies. b) Resulting linewidth $\Delta\nu$ (black dots) and contrast $C$ (blue triangles) depending on the modulation depth. c) Resulting slope $m$ (black dots) and noise $\sigma$ (blue triangles) depending on the modulation depth. d) Based on the measured slope and noise, the sensitivity $\eta_B$ (black dots) dependence on the modulation frequency is shown.

## S10 Double split-ring microwave resonator

As described in the main text, we used a double-split ring resonator to provide a microwave field necessary for the electron spin resonance measurements, which can be seen in Figure S12a. It consists of two circular spit-rings capacitively coupled to the feed strip line via a coupling gap. The geometrical parameters of the structure are summarized in Table S1 with corresponding scanning electron microscope images shown in Figure S13.

The resonance frequency of the empty resonator on the Rogers RO3010 substrate (thickness 0.64 mm, copper metallization 20 μm, dielectric constant $\epsilon = 10.2$, $\tan\delta = 2.2 \times 10^{-3}$) was calculated to be $f_{res} = 3.073$ GHz. The resonator was modeled to be undercoupled having an unloaded Q-value of $Q_0 = 147$ and a coupling coefficient of $\beta = 0.5$. The resonance frequency $f_{res}$ and the coupling coefficient $\beta$ of the structure are adjustable by positioning a thin metal plate above the outer split ring and a dielectric plate over the coupling gap, respectively. Figure S14 shows the simulation of $\beta$ dependent on the position of the coupler (Rogers RO3010 with dimensions of $0.6 \times 1.6 \times 0.3$ mm$^3$) and the resulting $f_{res}$ dependent on the position of the tuner (copper plate with $3.0 \times 1.2 \times 0.02$ mm$^3$) on the outer split-ring.

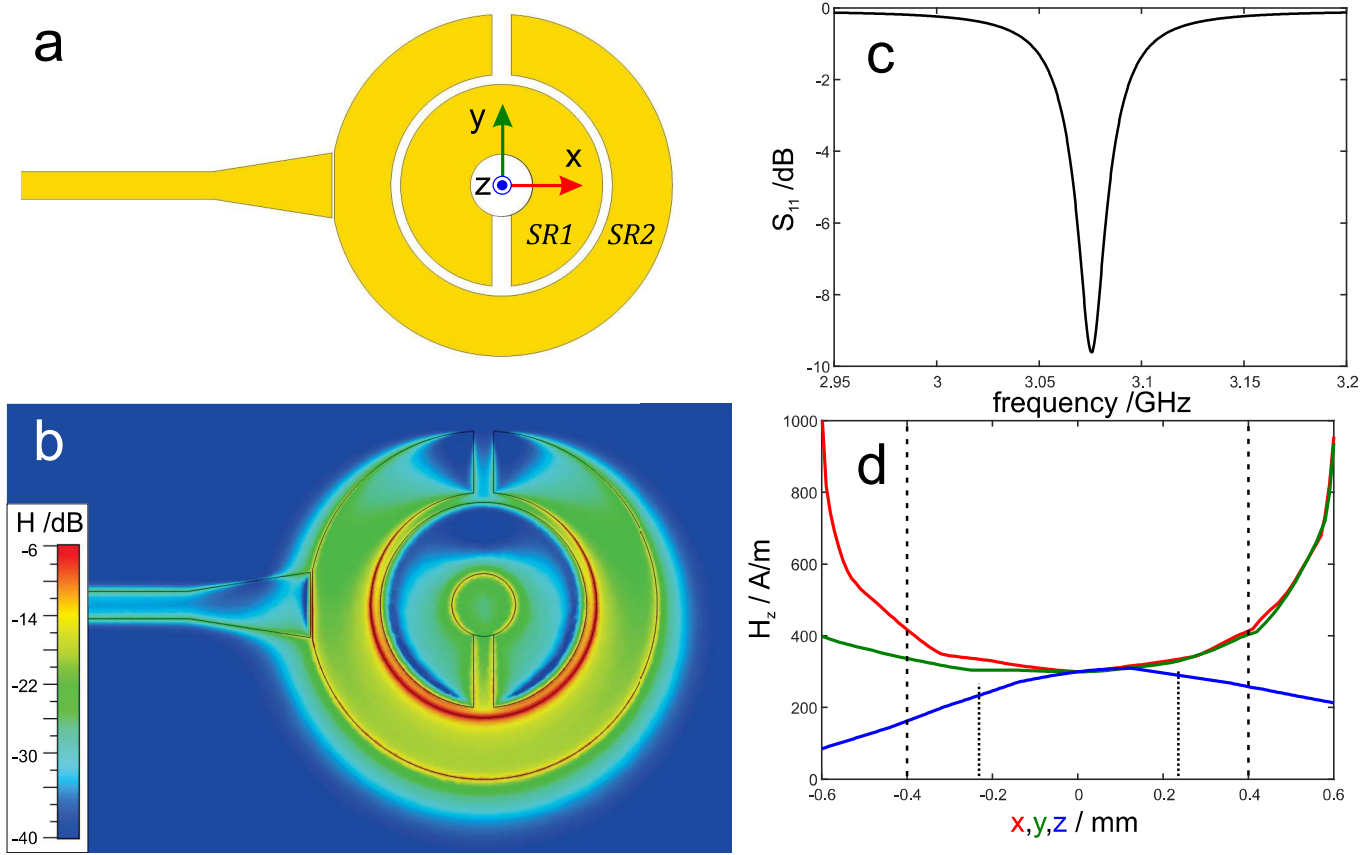

Figure S12: **Design and simulation of the double split-ring resonator.** a) The double split-ring resonator with coordinate system indicated. The geometrical parameters are summarized in Table S1. b) Contour plot of the normalized microwave magnetic field magnitude at $z = 0\,\mathrm{mm}$ in logarithmic scale was calculated using CST Microwave Studio®. c) Calculated $S_{11}$ curve for an empty resonator. d) The amplitude of $H_z$ magnetic microwave field component in dependence on $x$, $y$, and $z$ as indicated in a) for a critically coupled structure. The dashed and dotted lines indicate the sample size of $0.8 \times 0.8 \times 0.5\,\mathrm{mm}^3$.

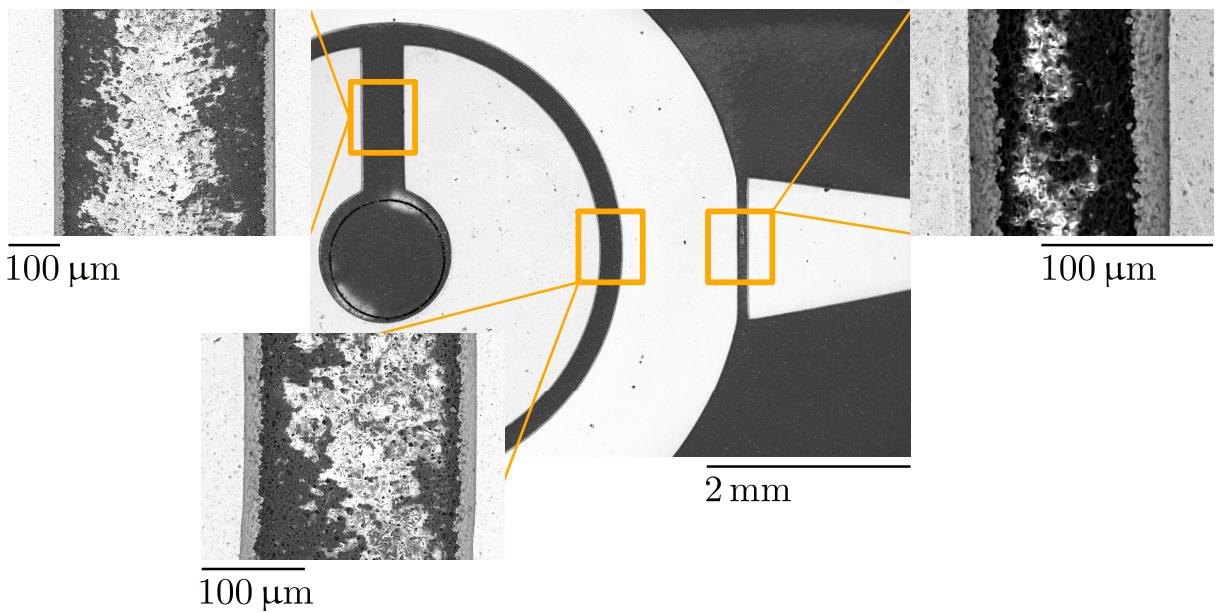

Figure S13: **Scanning electron microscope (SEM) image of the double split-ring resonator.** Images of the fabricated PCB were recorded by a SEM (Hitachi TM 3000 with energy of 15 kV). The width of the coupling gap was measured to be $(120.7 \pm 5.5)\,\mathrm{\mu m}$, the width of the gap between the inner and outer ring to be $(226.7 \pm 3.8)\,\mathrm{\mu m}$, and the width of the inner gap to be $(430.0 \pm 2.0)\,\mathrm{\mu m}$. The data was obtained by analyzing three fabricated PCBs, from which the values of mean and standard deviation were calculated.

Table S1: Geometrical parameters of the double split-ring resonator.

| Parameter | Value (mm) |
|---|---|
| $ID_{SR1}$ | 1.3 |
| $OD_{SR1}$ | 4.2 |
| $ID_{SR2}$ | 4.6 |
| $OD_{SR2}$ | 7.1 |
| Gap width (SR1, SR2) | 0.4 |
| Gap width (coupling) | 0.1 |
| Gap length (coupling) | 1.45 |

Based on the simulations using CST Microwave Studio®, an amplitude $H_z$ of the linearly polarized magnetic field along the $z$-axis of $310\,\text{A}\,\text{m}^{-1}$ was calculated at the resonator center for the critically coupled structure and 1 W incident microwave power. Over the sample volume of $0.8 \times 0.8 \times 0.5\,\text{mm}^3$ an average amplitude of the $z$-component of the magnetic field with $H_z = 335\,\text{A}\,\text{m}^{-1}$ was simulated for 1 W, see Figure S12d and Figure S15.

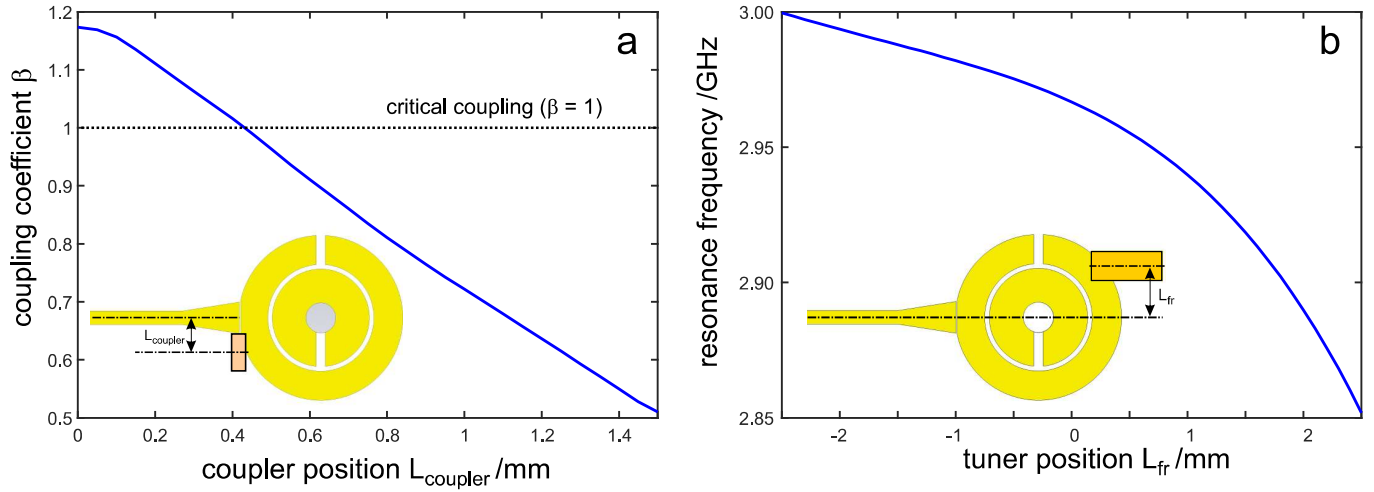

Figure S14: **Adjustments of the properties of the double split-ring resonator.** a) Adjustment of the coupling parameter. Calculated coupling coefficient in dependence on coupler (Rogers RO3010, $0.6 \times 1.6 \times 0.3\,\text{mm}^3$) position above the coupling gap of 0.1 mm. b) Adjustment of the resonance frequency. Calculated resonance frequency in dependence on the position of the frequency tuner (copper, $3.0 \times 1.2 \times 0.02\,\text{mm}^3$), which is placed on top of the surface of the outer split-ring.

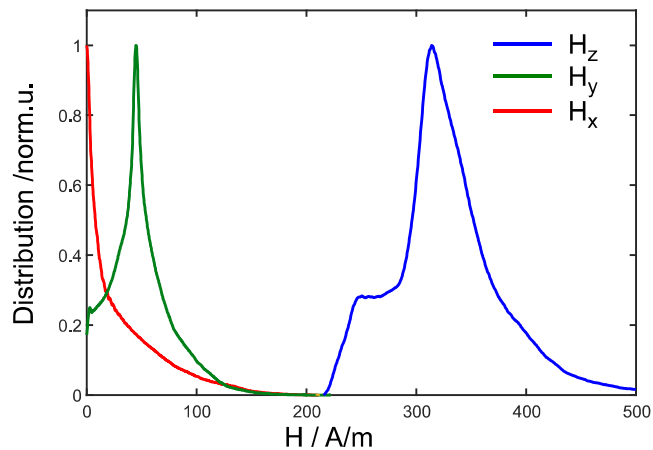

Figure S15: **Simulative result of the field distribution of the double split-ring resonator.** Amplitude distributions of the microwave magnetic field components $H_x$, $H_y$, and $H_z$ over the sample volume of $0.8 \times 0.8 \times 0.5\,\text{mm}^3$.

The fabricated structure is shown in Figure S16a. Figure S16b depicts the reflection curve $S_{11}$ of the resonator recorded by a network analyzer (Hewlett Packard 8753D) in the assembled NV sensor. The experimental curve was analyzed using a general expression for the reflection coefficient

$$S_{11} = 10 \cdot \log_{10}\left(|\Gamma|^2\right), \tag{S7}$$

where the reflection coefficient $\Gamma$ is given by

$$\Gamma = \frac{1 - \beta - i\xi}{1 + \beta - i\xi}, \tag{S8}$$

where $\beta$ is the coupling coefficient and $\xi = 2Q_0\left(\frac{f}{f_{\text{res}}} - \frac{f_{\text{res}}}{f}\right)$ an offset parameter. The numerical fit yields an unloaded Q-value of $Q_0 = 117.12$ and a coupling coefficient of $\beta = 0.3$, which are in a good agreement with the calculated values. The small deviation between calculated and experimental values is explained by the roughness of the substrate surface, imperfections of metallization from the fabrication process as well as the discrepancy of real dielectric parameters of substrate material, which are not considered in the simulation. The experimentally determined resonator bandwidth of $\Delta f = (1 + \beta) \cdot \frac{f_{\text{res}}}{Q_0} = 32.26\,\text{MHz}$ is in good agreement with predicted value of $30\,\text{MHz}$.

Figure S16c depicts the measured contrast of the optically detected electron spin resonance. For these measurements, the applied offset magnetic field was varied resulting in a change of the resonance frequency of the selected electron spin transition. As the spin resonance signal described by the contrast $C$ depends on the applied microwave field $B_1$, we compared the measured contrast with the induced $B_1$ by the following approximation

$$C \propto B_1 \propto \sqrt{\frac{P_{\text{abs}}}{P_0}} = \sqrt{1 - \frac{P_{\text{refl}}}{P_0}} = \sqrt{1 - |\Gamma|^2} = \sqrt{1 - 10^{S_{11}/10}}. \tag{S9}$$

The curves of $C$ and $B_1$ in Figure S16c show a similar behavior, which can be attributed to the characteristics of the double split-ring resonator.

We further analyzed the resonator structure by considering the extracted ODMR linewidths for different microwave powers $P_{\text{MW}}$, see Figure S16d. These data corresponds to the second optimization step regarding the sensors performance, where we evaluated the magnetic field sensitivity with respect to the applied microwave power, see main text. Typically, the NV linewidth broadens for an increased microwave field strength compared to the natural linewidth described by the inhomogeneously broadened linewidth $\Delta\nu_{\text{inh}}$ in a low power regime. [1] We therefore considered the measured ODMR linewidth to be the sum of $\Delta\nu_{\text{inh}}$ and a power induced linewidth $\Delta\nu_{\text{power}}$. This is in accordance with the linewidth model for NV ensembles described by Jensen et al.. [2] We used this model to fit the measured linewidth by using the following equation

$$\Delta\nu = \Delta\nu_{\text{inh}} + \Delta\nu_{\text{power}} = \frac{\gamma_2^*}{\pi} + \sqrt{\left(\frac{\gamma_2^{\text{eff}}}{\pi}\right)^2 + \frac{4\gamma_1^{\text{eff}}}{\gamma_2^{\text{eff}}} \cdot \left(\frac{\Omega_R}{2\pi}\right)^2}, \tag{S10}$$

where $\gamma_2^*$ is the inhomogeneous dephasing rate with $\gamma_2^* = 1/T_2^*$, $\gamma_1^{\text{eff}}$ and $\gamma_2^{\text{eff}}$ are the effective relaxation rates described by $\gamma_1^{\text{eff}} = 1/T_1 + \Gamma_P$ and by $\gamma_2^{\text{eff}} = 1/T_2 + \Gamma_P/2$ with $T_1$ the longitudinal and $T_2$ the transverse relaxation times, $\Gamma_P$ the optical pumping rate, and $\Omega_R$ the Rabi frequency in units of rad/s. The Rabi frequency is directly proportional to the microwave field, i.e. $\Omega_R \propto B_1 \propto \sqrt{P_{\text{MW}}}$. For the optical pumping rate, we considered the following relationship

$$\Gamma_P = \sigma \frac{I_{\text{opt}}}{E_{\text{ph}}}, \tag{S11}$$

where $\sigma$ is the absorption cross-section with $0.31 \times 10^{-16}\,\text{cm}^2$, [3] $I_{\text{opt}}$ the optical excitation intensity, and $E_{\text{ph}}$ the photon energy of the $521.9\,\text{nm}$ laser wavelength with $E_{\text{ph}} \approx 2.38\,\text{eV}$. We estimated an average optical power at the output of the GRIN lens of $18.3\,\text{mW}$ with a diameter of $75.2\,\mu\text{m}$ for a laser output power of $23.0\,\text{mW}$ due to losses in the optical path. We used relaxation times of $T_1 = 5.89\,\text{ms}$ and $T_2 = 96.49\,\mu s$, which have been evaluated for this diamond on a confocal microscope setup. For an NV concentration of 0.4 ppm, we calculated an average intensity along the optical path over diamonds thickness of $I_{opt} \approx 390.19\,W\,cm^{-2}$ resulting in a pumping rate of $\Gamma_P \approx 31.78\,kHz$. The fit of the model from Equation S10 revealed an inhomogeneously broadened linewidth of $\Delta\nu_{inh} = (68.67 \pm 2.99)\,kHz$, which is in good agreement with the minimal measured linewidth of $(68.79 \pm 3.30)\,kHz$, see Figure S9b. Concerning the power broadening we extracted $\Delta\nu_{power}$ to be $86.6\,kHz$ for a microwave input power of $1\,mW$, which is also in good agreement with the measured linewidth.

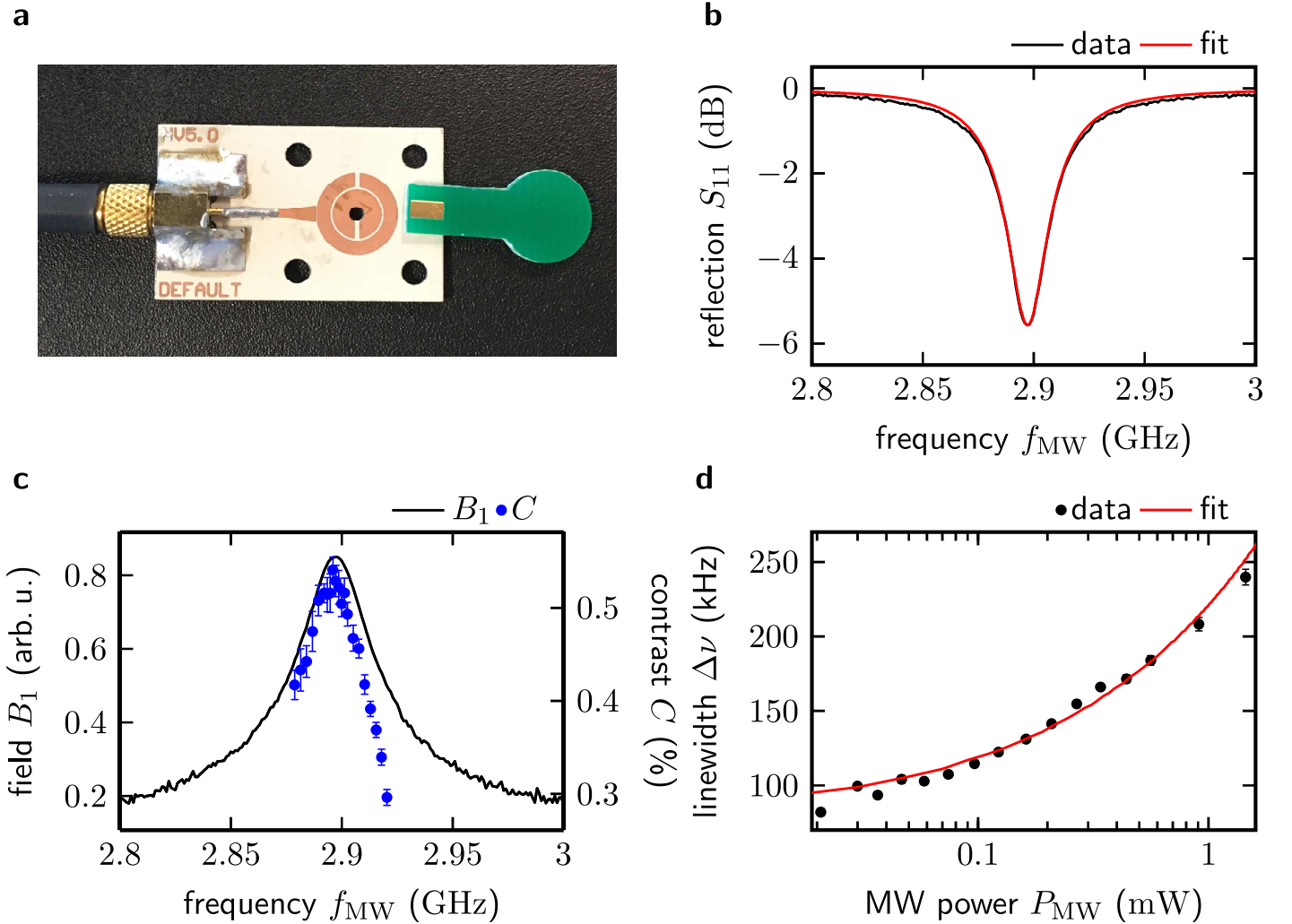

Figure S16: **Microwave double-split ring resonator.** a) Photo of the double split-ring resonator with the additional metallized plate to tune the resonance frequency. b) The $S_{11}$ coefficient data (black curve) measured by a network analyzer (Hewlett Packard 8753D) was used to determine the quality factor $Q_0$ and the coupling property of the microwave resonator by fitting the data. From a Lorentzian fit, we extracted a resonance dip of $(-5.44 \pm 0.01)\,dB$ and a width of $(32.25 \pm 0.11)\,MHz$ (red curve) resulting in $Q_0 = 117.12$ and $\beta = 0.3$. The structure was modeled to be undercoupled without a dielectric coupler resulting in $Q_0 = 185$ and $\beta = 0.31$. c) Measured contrast $C$ (blue dots) for different strengths of the offset magnetic field generated by the tapered Helmholtz coils. For the experiments, the resonator was tuned to a resonance frequency $f_{res} = 2.898\,GHz$ by adjusting the position of a metallized plate on the resonator structure that varied the capacitance of the split ring resonator. A microwave power of $36\,\mu W$ was used. Based on the measured $S_{11}$ data shown in b) the $B_1$ field strength (black curve) according to Equation S9 is added to this graph. d) Measured linewidth $\Delta\nu$ (black dots) depending on the applied microwave power $P_{MW}$ (data from Figure S9b). The values of the linewidth were extracted from fitting the recorded spectra using the Lorentzian function of Equation S6. We applied the model of the NV linewidth by Jensen *et al.*, see Equation S10, to the data (red curve), from which the Rabi frequency $\Omega_R$ can be derived using $\Omega_R \propto \sqrt{P_{MW}}$.

## S11  Magnetic test field

Besides the determination of the magnetic sensitivity by taking the measured parameters of slope and noise of the zero-crossing point (Figure 5), we further evaluated the sensor performance under the application of an additional magnetic field $B_{\text{test}}$. A copper wire was wound around the pair of Helmholtz coils and we applied a DC field to the wire by an arbitrary waveform generator (Agilent 33220A). The generated flux density at the center of the Helmholtz coil pair (with a coil radius $r$ and $N$ turns on each side) corresponds to the superposition of two circular currents and can be calculated according to Biot-Savarts law by [4]

$$B_{\text{test}} = \frac{8}{\sqrt{125}} \frac{\mu_0 N}{r} \frac{U}{R}, \tag{S12}$$

where $\mu_0$ is the vacuum permeability with $\mu_0 = 4\pi \times 10^{-7}\,\text{N}\,\text{A}^{-2}$, $U$ the applied voltage, and $R$ the resistance of the coil wires. For a generated output with an amplitude of $10\,\text{mV}$ and a resistance of $229.8\,\Omega$, we expected a magnetic field of $B_{\text{test}} \approx 6.52\,\text{nT}$ at the center of the Helmholtz coils. To experimentally verify this induced magnetic field, we applied a DC field with voltage of $500\,\text{mV}$ and recorded the ODMR spectrum. A shift of the resonance frequency of $9.34\,\text{kHz}$ was observed for the NV orientation with the largest splitting, which was selected for this analysis. This resonance shift corresponds to an additional magnetic field of approximately $333\,\text{nT}$. Here, we neglected temperature effects as the measurements haven been carried out in close succession. Assuming a linear scaling of the field strength and the applied voltage, the projected strength of the test field would be $6.66\,\text{nT}$ on the selected NV orientation for $10\,\text{mV}$. The calculated value from Equation S12 is in good agreement with the experimentally verified frequency shift observed in the ODMR measurement.

# References

[1] A. Dréau, M. Lesik, L. Rondin, P. Spinicelli, O. Arcizet, J.-F. Roch, V. Jacques, *Physical Review B* **2011**, *84* 195204.

[2] K. Jensen, V. M. Acosta, A. Jarmola, D. Budker, *Physical Review B* **2013**, *87*, 1 014115.

[3] T.-L. Wee, Y.-K. Tzeng, C.-C. Han, H.-C. Chang, W. Fann, J.-H. Hsu, K.-M. Chen, Y.-C. Yu, *The Journal of Physical Chemistry A* **2007**, *111*, 38 9379 .

[4] V. E. Baranova, P. F. Baranov, In *2014 Dynamics of Systems, Mechanisms and Machines (Dynamics)*. **2014** 1–4.